\documentclass[fleqn,usenatbib]{mnras}

\usepackage[T1]{fontenc}
\usepackage{ae,aecompl}

\usepackage[utf8]{inputenc}

\usepackage{float}
\usepackage{amsmath,amssymb,amsfonts}
\usepackage{mathrsfs}
\usepackage{wasysym}
\usepackage{fancyhdr}

\usepackage{graphicx}
\usepackage{caption}
\usepackage{color}
\usepackage{hyperref}

\usepackage{theorem}
\newtheorem{teo}{Theorem}[section]
\newtheorem{corollary}{Corollary}[section]

\definecolor{blue}{rgb}{0,0,1}
\definecolor{green}{rgb}{0,0.65,0.5}
\definecolor{verde}{rgb}{0.,.5,0.4}
\definecolor{marron}{rgb}{0.7,0.2,0.1}
\definecolor{red}{rgb}{1,0,0}
\definecolor{vio}{rgb}{0.66,0,1}
\definecolor{ama}{rgb}{1,1,0}
\definecolor{veroscuro}{rgb}{0.3,0.36,0.33}

\title[Efficient gravitational lens optical scalars...]{Efficient gravitational lens optical scalars 
	calculation of black holes with angular momentum}

\author[E.F.Boero and O.M.Moreschi]{
	Ezequiel F. Boero,$^{1}$\thanks{E-mail:boero@famaf.unc.edu.ar}
	Osvaldo M. Moreschi,$^{2,3}$\thanks{E-mail:moreschi@fis.uncor.edu}
	\\
	$^{1}$ Instituto de Astronom\'\i{}a Te\'{o}rica y Experimental (IATE), CONICET,
	Observatorio Astron\'{o}mico de Córdoba,\\
	Laprida 854, (X5000BGR) C\'{o}rdoba, Argentina.\\
	$^{2}$Facultad de Matem\'{a}tica, Astronom\'\i{}a, F\'\i{}sica y Computaci\'{o}n (FaMAF),
	Universidad Nacional de C\'{o}rdoba, \\ 
	Medina Allende S/N, X5000HUA, Córdoba, Argentina.  \\
	$^{3}$Instituto de F\'\i{}sica Enrique Gaviola, IFEG, CONICET,Medina Allende S/N, X5000HUA, Córdoba, Argentina. 
}

\date{Accepted XXX. Received YYY; in original form ZZZ}

\pubyear{2019}

\begin{document}
	\label{firstpage}
	\pagerange{\pageref{firstpage}--\pageref{lastpage}}
	\maketitle
	
\begin{abstract}
We provide new very simple and compact expressions for the efficient 
calculation of gravitational lens optical scalars for Kerr spacetime
which are exact along any null geodesic.
These new results  are obtained recurring 
to well known results on geodesic motion that exploit obvious and hidden symmetries 
of Kerr spacetime and contrast with the rather long and cumbersome expressions 
previously reported in the literature, providing
a helpful improvement for the sake of an efficient integration of the geodesic 
deviation equation on Kerr geometry.	
We also introduce a prescription for the observer frame which captures a new 
notion of \emph{center of the black hole} which can be used for any position 
of the observer, including those near the black hole.
We compare the efficient calculation of weak lens optical scalars
with the exact equations; finding an excellent agreement.
\end{abstract}


\section{Introduction}

Black holes without intrinsic angular momentum are well represented by
the Schwarzschild vacuum static spherically symmetric solution.
The weak lensing of these geometries have been extensively studied;
in particular in terms of approximations based on formalisms
using scalars, and the projection of mass distribution on the
plane of the thin lens model\cite{Schneider92, petters2001singularity}.
It is generally believed that
more general lenses distributions can be discussed 
in the weak field regime by linear combination of the effect produced
by the Schwarzschild solution;
however it is known that these type of approximations miss general relativistic
interesting systems\cite{Gallo11,Boero:2016nrd} over flat or cosmological
backgrounds.
In particular these models do not describe the Kerr gravitation lens
as we mention below.
The presence of angular momentum delivers new difficulties 
for the program of dealing 
with a lens equation in terms of expressions involving a notion of bending angle;
which has been emphasized in several works, for example \cite{Aazami:2011tu, Aazami:2011tw}
and also \cite{Renzini:2017bqg, Werner:2007vu, Sereno:2006ss}.
Furthermore, the lens optical scalars can not be obtained from a scalar function such as the
lens potential employed in the standard formalism \cite{Schneider92, Schneider06}.

When the black holes have angular momentum, their geometries are adequately
represented by the Kerr vacuum stationary axisymmetric solutions.
In the case of supermassive black holes with angular momentum;
the surrounding matter could be considered as a 
first order test field contribution on
Kerr spacetime.
But in these situations the methods on gravitational lens
based on the projection of mass distribution on the 
plane of a thin lens do not work.
This is mainly due to the fact that the component of the Weyl
curvature in the frame adapted to the observed photon null geodesic,
reveals the expected spin 2 gravitational nature; which in
the spherically symmetric case can be circumvented. 
This point is particularly important in works calculating the image
of a black hole with angular momentum, as are those 
related to the publications of the Event Horizon Telescope 
Collaboration\cite{Akiyama:2019cqa,Akiyama:2019brx,Akiyama:2019sww,
	Akiyama:2019bqs,Akiyama:2019fyp,Akiyama:2019eap}.

The Kerr metric is not only important from the formal point of view, for
being the vacuum solution for stationary black holes with angular momentum;
but also because any isolated black hole with angular momentum, independently 
of its origin, is expected to decay rapidly to the geometry described by 
the Kerr vacuum solution of the Hilbert-Einstein equations\cite{Carter:1971zc}.
For this reason the Kerr metric has captured the
interest of the community to extend the lens formalism for the study of weak 
lensing effects in black holes with angular momentum 
and has motivated the appearance of 
works\cite{Cunningham_1973ApJ, Bray1986, Asada:2000vn, Asada_2003, 
	Sereno:2006ss, Werner:2007vu, Gyulchev:2008ff, Iyer:2009wa, Kraniotis:2010gx, 
	Renzini:2017bqg}
on the subject.

In this work we present a new result which is a
convenient expressions to calculate exact gravitational 
lens optical scalars in the Kerr spacetime.
From the mathematical point of view, one is interested in the problem
of solving the \emph{null geodesic equation}, of the observed photons,
together with the \emph{geodesic deviation equation}
for arbitrary position of the observer, instead to rely on approximations based on 
linearized gravity.
For a vacuum spacetime, as Kerr, the geodesic deviation equation 
can be expressed  in terms of the component $\Psi_0$ of the Weyl curvature,
with respect to the geodesic frame using the GHP notation\cite{Geroch73};
for which we have obtained a very compact exact expression.
We explain below how this component is calculated.

We also present a detailed discussion of the observational setting that an stationary observer
should  use, 
which is a convenient way to describe the formation of images and the distortion effects 
by means of the optical scalar.
Our work includes a construction of a frame for the observer which is based in the unique null geodesic
passing by the observer spacetime event, belonging to the 
\emph{center of mass null geodesic congruence}\cite{Arganiaraz19a}.
We claim that such choice is very helpful since it provides a clear way to identify a notion of `center' 
for the black-hole in the sky of the observer. 
This is particularly useful when comparing observations of systems with 
diverse angular momentum and different inclination angles of observation.

We apply our construction to the case of a supermassive black hole where the
mass and angular momentum are taken from that found in galaxy M87;
although we use an arbitrary inclination of the angular momentum
with respect to the line of sight.

This article is organized as follows.
For completeness, in section \ref{sec:GeodDev+WeakLens} we recall the
set of equations that are needed for the exact calculation of the lens effects.
In section \ref{sec:standardweaklensapprox} we recall the 
standard weak lens approximations.
The geometry of the Kerr spacetime is presented in section \ref{sec:kerr}.
The calculation of the constants of motion in terms of the observed
angles is done in section \ref{sec:Integr+Const}
as well as the particular construction of the observational frame.
In section \ref{sec:lensing-Kerr} we present our main results of the calculation
of the Weyl curvature component $\Psi_0$.
As an application  we show in section \ref{sec:numerical}
the  comparison of the exact numerical calculation of
the shear with the first order weak lens approximation.
We also include few appendices 
for completeness of the presentation.

\section{Exact gravitational lens optical scalars}\label{sec:GeodDev+WeakLens}

\subsection{Null geodesics and the geodesic deviation equation}\label{subsec:Geod+Dev}

Weak lensing effects deal mainly with the deformation of images of objects that
are behind some distribution of matter or geometry, the lens.
Typically one encodes this deformation in the optical scalars that describe
expansion, shear and twist of the images.
All these are related to the behavior of null geodesics
in the vicinity of a particular ray. From the mathematical point of view,
then, they are described in terms of the \emph{geodesic equations}
and the \emph{geodesic deviation equations} that we recall below.

By \emph{exact gravitational lens optical scalars} we mean the exact calculations of the effects described above.
This should not be confused with the usual approximations that only take into account
the first order effects of the curvature of the lens over some background.

Let us then consider an infinitesimal small bundle from a congruence of null geodesics 
starting at the source and reaching the position of the observer.
The affinely parametrized tangent vector field $\ell^a$ to the central geodesic of 
such bundle satisfies the \emph{geodesic equations}
\begin{equation}\label{eq:geod+ell}
\ell^a \nabla_a\ell^b = 0;
\end{equation}
where we are using the standard abstract index notation with Latin
letters $a,b,c,...$

Following the notation in \cite{Boero:2016nrd},
the freedom concerning the specifications of the affine parameter 
$\lambda$ is set by taking it to be zero at the position of the 
observer and scaled in such way that if $v^a$ denotes the $4$-velocity of the 
observer it satisfies
\begin{equation}\label{eq:normalization+lambda}
\ell^a v_a = 1.
\end{equation}
Local distortions of the cross section of the bundle 
can be described by means of the \emph{geodesic deviation vector}, denoted by $\varsigma^a$, 
which has spacelike character, namely $\varsigma^a \varsigma_a < 0$ 
(with our choice of metric signature $(+,-,-,-)$) and satisfies
\begin{equation}\label{eq:Lie+prop+devVec}
\mathscr{L}_{\ell} \; \varsigma^{a} = 0;
\end{equation}
i.e. it is Lie propagated along the null congruence. 
From equations (\ref{eq:geod+ell}) and (\ref{eq:Lie+prop+devVec}) one deduces that 
it also satisfies the \emph{geodesic deviation equation}
\begin{equation}\label{eq:desv+of+geod+1}
\ell^a \nabla_a \left( \ell^b \nabla_b \varsigma^d \right) = 
R_{abc}^{\;\;\;\;\;d}\ell^a \varsigma^b \ell^c;	
\end{equation}
where as usual $R_{abc}^{\;\;\;\;\;d}$ denotes the Riemann tensor of the spacetime.

For future calculations along the article, we will employ these equations expressed 
in terms of the components of a null tetrad $(\ell^a, m^a, \bar{m}^a, n^a)$; 
where $m^a$ and $\bar{m}^a$ is a pair of complex conjugated vectors
and $n^a$ an additional real null vector.
The two complex vectors $m^a$ and $\bar{m}^a$ will be chosen to be parallel propagated 
along the geodesic. 
The choice of tetrad is related to some nice algebraic 
properties\footnote{
Let us note that this will have as a consequence that the spin coefficient\cite{Geroch73}
$\epsilon$ vanishes.
}
which will become convenient in the calculations of section \ref{sec:lensing-Kerr}.

The fourth tetrad vector, $n^a$ is picked up in such way that the tetrad satisfies the 
relations:
\begin{align}
\ell^a n_a =& ~1, \\
m^a \bar{m}_a =& -1;
\end{align}
with all the other possible contractions being zero.

The geodesic deviation vector then can be expressed as
\begin{equation}
\varsigma^a = \varsigma \bar{m}^a + \bar{\varsigma} m^a;
\end{equation}
and it is convenient to define the two component array $\mathcal{X}$ by:
\begin{equation}
\mathcal{X} \equiv \begin{pmatrix}
\varsigma \\
\bar{\varsigma}
\end{pmatrix};
\end{equation}
since, following standard procedures\cite{Gallo11,Boero:2016nrd}
it satisfies the following equation
\begin{equation}\label{eq:dev+geoII}
\ell\left(\ell\left(\mathcal{X}\right)\right) = 
- Q\mathcal{X} \; ;
\end{equation}
which is equivalent to the geodesic deviation equation (\ref{eq:desv+of+geod+1}),
where the matrix $Q$ is 
\begin{equation}\label{eq:matrix+Q}
Q = 
\begin{pmatrix}
\Phi_{00} & \Psi_{0} \\
\bar{\Psi}_{0} & \Phi_{00}
\end{pmatrix} ;
\end{equation}
whose elements are the curvature scalars\cite{Geroch73}:
\begin{equation}
\Phi_{00} = -\frac{1}{2}R_{ab} \, \ell^{a} \, \ell^{b}  \, ,
\end{equation}
and
\begin{equation}\label{eq:Psi0+def}
\Psi_{0} = C_{abcd} \, \ell^{a} \, m^{b} \, \ell^{c} \, m^{d} \, .
\end{equation}
Here $R_{ab}$ and $C_{abcd}$ denote the Ricci and the Weyl tensor respectively, 
and as usual we denote with
$\,\bar{}\,$ the complex conjugation.

\subsection{Exact geodesic deviation equations as a first order system}\label{subs:gdefirst}

Any procedure to find exact solutions of the equation (\ref{eq:dev+geoII}) requires
to solve simultaneously the geodesic equation (\ref{eq:geod+ell}) in order to provide 
the path of integration.
Since the calculations are normally carried out by numerical techniques,
it is convenient to have at hand a representation of the geodesic deviation
equations as a first order system. 
This is recalled next.

Defining\cite{Gallo11} $\mathcal{V}$ to be
\begin{equation}
\mathcal{V} \equiv \frac{d\mathcal{X}}{d\lambda} ,
\end{equation}
and
\begin{equation}
\mathbf{X} \equiv \begin{pmatrix}
\mathcal{X} \\
\mathcal{V}
\end{pmatrix};
\end{equation}
one obtains
\begin{equation}\label{eq:dotx}
\ell(\mathbf{X}) = \frac{d\mathbf{X}}{d\lambda}
=
\left(
\begin{array}{c}
\mathcal{V} \\
-Q \,\mathcal{X}
\end{array}
\right) 
= A \, \mathbf{X}
;
\end{equation}
with
\begin{equation}
A \equiv
\left(
\begin{array}{cc}
0 & \mathbb{I} \\
-Q & 0
\end{array}
\right) .
\end{equation}
In terms of the components, one can also express these equations as:
\begin{align}
\dot{\varsigma} =& v_{\varsigma}, \\
\dot{v}_{\varsigma} =& - \Phi_{00} \varsigma - \Psi_0 \bar{\varsigma}, \\
\dot{\bar{\varsigma}} =& \bar{v}_{\varsigma}, \\
\dot{\bar{v}}_{\varsigma} =& - \bar{\Psi}_0 \varsigma -  \Phi_{00} \bar{\varsigma}.
\end{align}

\subsection{Exact gravitational lens optical scalars}\label{sec:weak+lensing}

Although it is customary in any discussion of weak lensing to assume
some kind of approximation either in the calculation, or in the nature
of the spacetime, or both; we here would like to rescue the notion
of exact optical scalars, to which approximations
can be applied.

The notion of the optical scalars comes from the comparison between the
non-lensing situation (i.e.: no gravity, and so flat geometry),
with the lensing observation due to the curved nature of the spacetime.

The language that one uses is the following.

\begin{figure}
	\centering
	\includegraphics[clip,width=0.4\textwidth]{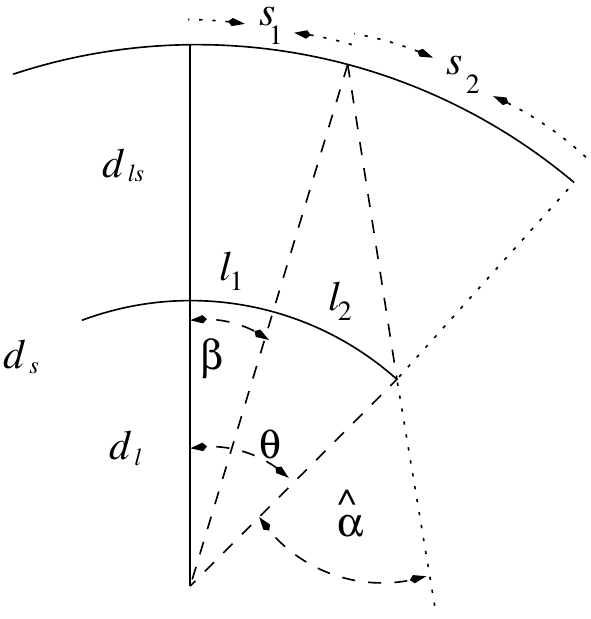}
	\caption{This graph shows the basic and familiar angular variables in terms of a simple
		flat background geometry.
		The letter $s$ denote sources, the letter $l$ denotes lens and the observer is
		assumed to be situated at the apex of the rays.}\label{fig:lens1}
\end{figure}

If a small portion of the source is observed to have an angular size
characterized by a vector $\delta \theta^a$ and $\delta \beta^a$ is the 
angular size that one would observe 
in the `absence' of the lens, then we 
can write the linear relation
\begin{eqnarray}
\delta \beta^a = \mathcal{A}^a_{\;\,b} \delta \theta^b;
\end{eqnarray}
where the matrix $\mathcal{A}^a_{\;\,b}$ is the differential map between
the two sets of angles in the sphere of observed directions.
In figure (\ref{fig:lens1}) below we present an sketch and define the quantities that 
are normally used in this discussions.

The general form of the matrix can be expressed in terms of the
\emph{optical scalars} $(\kappa, \gamma_1, \gamma_2, \omega)$ by:
\begin{equation}
\mathcal{A}^a_{\;\,b} = \begin{pmatrix}
1 - \kappa - \gamma_1 & -\gamma_2 - \omega \\
-\gamma_2 + \omega    & 1 - \kappa + \gamma_1
\end{pmatrix};
\end{equation}
where $\kappa$ denotes the expansion,
$\gamma_1$ and $\gamma_2$ are the components of the complex
shear $\gamma_1 + i \gamma_2$, and $\omega$ has the information of the twist.

Let us notice that when
an observer sees an angular size
$\delta \theta^a$, which, after integrating the previous geodesic
and deviation vector equations, 
corresponds to a deviation vector 
$\varsigma_s^a$ at the source; 
then, since $\beta^a = \varsigma_s^a/\lambda_s$,
with $\lambda_s \equiv \lambda_\text{observer} -  \lambda_\text{source}$,
the matrix of the optical scalars relates these quantities as:
\begin{equation}
\varsigma_s^a = \mathcal{A}^a_{\;\,b} (\lambda_s  \delta \theta^b )
.
\end{equation}

The magnification of the observed images is given by the inverse of the determinant;
which is:
\begin{equation}
\mu = \frac{1}{\left(1 - \kappa\right)^2 - \left(\gamma_1^2 + \gamma_2^2\right) + \omega^2} 
.
\end{equation}

\section{Standard weak lensing approximations}\label{sec:standardweaklensapprox}

For the sake of completeness we recall in this section the standard weak lensing equations
that we will use along with other related approximations.

\subsection{General expression for the optical scalars in the linear approximation}

The standard weak lens approximation involves the assumption that only linear curvature
effects are considered in the calculations.
In this case one can prove that there are no twist effects, so that
the weak lens matrix can be expressed as:
\begin{equation}
\mathcal{A}^a_{\;\,b} = \begin{pmatrix}
1 - \kappa - \gamma_1 & -\gamma_2 \\
-\gamma_2     & 1 - \kappa + \gamma_1
\end{pmatrix}.
\end{equation}
Correspondingly the magnification is expressed as:
\begin{equation}
\mu = \frac{1}{\left(1 - \kappa\right)^2 - \left(\gamma_1^2 + \gamma_2^2\right)} .
\end{equation}

Let us note that although in this framework one is considering linear contributions
coming from the curvature, it does not mean that large distortions 
such as arcs and rings can not take place.
In fact, such impressive distortions can be well described in these approximations
and are usually referred to as strong lensing effects which might be confusing 
since even though the distortions are strong, the effects are computed in a weak
field regime of the lens distribution.

Applying the techniques and notation of reference \cite{Gallo11} one can express
the convergence and shear in terms of the curvature scalar $\Phi_{00}$ and $\Psi_0$ 
of the lens, respectively in the following way:
\begin{equation}\label{eq:kappa+weak+lensing}
\begin{split}
\kappa =&
\frac{1}{\lambda_s} \int_{0}^{\lambda_s} \lambda\left(\lambda_s - \lambda \right)
\Phi_{00}(\lambda) \, d\lambda 
\\
=&
\int_{0}^{\lambda_s} \left(\frac{1}{ {\lambda}^2 }\int_{0}^{\lambda}
\Phi_{00}(\lambda') \, {\lambda'}^2 \, d\lambda' \right) d\lambda 
,
\end{split}
\end{equation}
\begin{equation}\label{eq:shear+weak+lensing}
\begin{split}
\gamma_{1} + i\gamma_{2} =& 
\frac{1}{\lambda_s} \int_{0}^{\lambda_s} \lambda\left(\lambda_s - \lambda \right)
\Psi_{0}(\lambda) \, d\lambda 
\\
=&
\int_{0}^{\lambda_s} \left(\frac{1}{ {\lambda}^2 }\int_{0}^{\lambda}
\Psi_{0}(\lambda') \, {\lambda'}^2 \, d\lambda' \right) d\lambda.
\end{split}
\end{equation}

It is important to remark that these expressions are manifestly gauge invariant, 
since one is considering departures from a flat background,
and the linear curvature is gauge invariant in this case.
Also one is considering contribution from the whole curvature and therefore from the 
spacelike components of the energy-momentum tensor as well.

\subsection{The standard thin lens approximation}

It is often the case that the lens can be assumed to be thin.
In those circumstances one can prove\cite{Gallo11}
that the optical scalars can be expressed by
\begin{equation}\label{eq:g1masig}
\begin{split}
\kappa = \frac{d_ld_{ls}}{d_s}\hat{\Phi}_{00},
\end{split}
\end{equation}
\begin{equation}
\begin{split}\label{eq:g1masig2}
\gamma_1+i\gamma_2 = \frac{d_ld_{ls}}{d_s}\hat{{\Psi}}_0,
\end{split}
\end{equation}
where
\begin{align}
\label{eq:integrated-ricci}
\hat{\Phi}_{00} &= \int^{d_s}_0\Phi_{00}d\lambda,\\
\label{eq:integrated-weyl}
\hat{\Psi}_{0} &= \int^{d_s}_0\Psi_{0}d\lambda,
\end{align}
are the projected curvature scalars along the line of sight, and
where $d_{l}$ denotes distance from the observer to the lens, $d_{{ls}}$ the distance from 
the lens to the source, and  $d_{s}$  the distance from the observer to the source.
(See figure \eqref{fig:lens1}.)

\subsection{The thin lens approximation in the cosmological context}

If the lens, which could be our Kerr black hole, is immersed in a cosmological background,
then one has to rethink all the calculations; since now the background is not
flat.
In reference \cite{Boero:2016nrd} we have extending this setting to the case in
which the background is a Robertson-Walker cosmology; so that
one can consider, for example
a lens that in its vicinity resembles the Kerr geometry, but that
asymptotically behaves as a Robertson-Walker spacetime.
In this setting one must assume that the typical size of the lens
is much smaller that the cosmological dimensions involved.
We have also allow in \cite{Boero:2016nrd} for the lens to be
moving with respect to the cosmological background. Then, for this
general case, we have shown that the expressions for the optical scalars are:
\begin{equation}\label{eq:kappa+thin+lensing}
\kappa = 
(1-\kappa_c)
\frac{1 + z_v}{1 + z_l} 
\frac{D_{A_{ls}} D_{A_{l}}}{D_{A_s}}
\int_{0}^{\lambda_s} 
\Phi_{00} \, d\lambda'
+ \kappa_c
,
\end{equation}
\begin{equation}\label{eq:shear+thin+lensing}
\gamma_{1} + i\gamma_2 = 
(1-\kappa_c)
\frac{1 + z_v}{1 + z_l}
\frac{D_{A_{ls}} D_{A_{l}}}{D_{A_s}}
\int_{0}^{\lambda_s} 
\Psi_{0} \, d\lambda';
\end{equation}
where $\kappa_c$ is the contribution for the convergence coming from the cosmological
background,
$z_v$ is the redshift of the lens,
$z_l$ is the cosmological redshift at the lens position,
$D_{A_{l}}$ denotes area distance from the observer to the lens,
$D_{A_{ls}}$ the area distance from the lens to the source, and
$D_{A_s}$ the area distance from the observer to the source.

\section{Kerr spacetime}\label{sec:kerr}
\subsection{The metric in Boyer-Lindquist coordinates}\label{subsec:Boyer-Lindquist}

The line element of the Kerr metric in the Boyer-Lindquist 
coordinate\cite{Boyer67} system 
$(t,r,\theta, \phi)$ is given by:
\begin{equation}\label{eq:Kerr-usualBL-form}
\begin{split}
ds^2 
=& \left( 1 - \varPhi\right) dt^2  
+ 2 \varPhi a \sin^2(\theta) dt d\phi 
- \frac{\Sigma}{\Delta} dr^2 
\\
& - \Sigma d\theta^2 
- \left( r^2 + a^2 + \varPhi a^2  \sin^2(\theta)\right) \sin^2(\theta)
d\phi^2
;
\end{split}
\end{equation}
where $M$ and $a$ are the mass and rotation parameter respectively, and the functions 
$\Sigma(r, \theta) $, $\Delta(r)$ and $\varPhi(r,\theta)$, are 
defined as
\begin{align}
\Sigma &= r^2 + a^2 \cos(\theta)^2, 
\label{eq:Sigma-def} \\
\Delta &= r^2 - 2rM + a^2 , \label{eq:Delta-def} 
\end{align}
and
\begin{equation}
\varPhi =\frac{2 M r}{\Sigma}  \label{eq:Phi-def} 
.
\end{equation}

\subsection{The repeated principal null directions}\label{sec:PNTetrad}

In our work we  benefit from the fact that the Kerr metric is
type D, meaning that it has two principal null geodesic congruences.
It is then possible to choose a null tetrad in which two of
the null vectors are parallel to these two principal null directions.
This  turn out to be useful when we later introduce a null tetrad 
adapted to the null geodesic that one considers in the 
weak lens analysis, in particular when dealing with  the deviation 
geodesic equation (\ref{eq:dev+geoII}).

Denoting by $\big( \tilde{\ell}, \tilde{m}, \bar{\tilde{m}}, \tilde{n}^a \big)$, 
we take $\tilde{\ell}^a$ as the real outgoing principal null vector,
with affine parametrization, while the other real null vector $\tilde{n}^a$
is chosen so that it is parallel to the second (in going) principal null direction. 
In terms of Boyer-Lindquist coordinates, a common choice for such a null tetrad is 
the following one \cite{Chandrasekhar:1985kt}:
\begin{align}
\tilde{\ell}^a &= \frac{r^2 + a^2}{\Delta}\partial_t^a + \partial_r^a + \frac{a}{\Delta}\partial_{\phi}^a, \\
\tilde{n}^a  &= \frac{r^2 + a^2}{2\Sigma}\partial_t^a - \frac{\Delta}{2\Sigma}\partial_r^a + \frac{a}{2\Sigma}\partial_{\phi}^a, \\
\tilde{m}^a &= \frac{ia\sin(\theta)}{\sqrt{2} \mathfrak{r}}\partial_t^a 
+ \frac{1}{\sqrt{2} \mathfrak{r}}
\bigg(
\partial_\theta^a +
\frac{i}{\sin(\theta)}\partial_\phi^a
\bigg)
;
\end{align}
where the complex function $\mathfrak{r} $ is defined as:
\begin{equation}\label{eq:rho_hat}
\mathfrak{r} = r + i a \cos(\theta)
,
\end{equation}
and $\partial_x \equiv \frac{\partial}{\partial x}$.
If we lower the indices the expressions become
\begin{align}
\begin{split}
\tilde{\ell}_a &= dt_a - \frac{\Sigma}{\Delta }dr_a - a \sin(\theta)^2 d\phi_a  , \label{eq:ell+form+princ}
\end{split}
\\
\begin{split}
\tilde{n}_a  &= \frac{\Delta}{2\Sigma} dt_a + \frac{1}{2} dr_a - \frac{a\Delta \sin(\theta)^2}{2\Sigma} d\phi_a ,
\end{split}
\\
\begin{split}
\tilde{m}_a &= \frac{ia\sin(\theta)}{\sqrt{2}\mathfrak{r}}dt_a 
- 
\frac{1}{\sqrt{2}\mathfrak{r}}
\bigg(
\Sigma d\theta_a 
+ i(r^2 + a^2)\sin(\theta) d\phi_a
\bigg)
;\label{eq:m+form+princ}
\end{split}
\end{align}
and one can easily corroborate the normalization conditions, namely
\begin{align}
\tilde{\ell}^a \tilde{n}_a &= 1; \\
\tilde{m}^a \bar{\tilde{m}}_a &= -1.
\end{align}
The metric then can be written as
\begin{equation}
g_{ab} = \tilde{\ell}_a \tilde{n}_b + \tilde{n}_a \tilde{\ell}_b - \tilde{m}_a \bar{\tilde{m}}_b -
\bar{\tilde{m}}_a \tilde{m}_b
.
\end{equation}

Since the chosen null tetrad includes both of the principal null directions,
the curvature is expressed in terms of a single scalar, namely  $\tilde{\Psi}_2$;
which is the Weyl null tetrad component of spin weight zero.
Then, 
the Weyl curvature scalar $\Psi_0$ of equation (\ref{eq:dev+geoII})
(in the null frame adapted to the photon trajectory)
must be 
proportional to this only non-vanishing Weyl scalar 
which is given by:
\begin{equation}
\tilde{\Psi}_2 = C_{abcd}\tilde{\ell}^a \tilde{m}^b \bar{\tilde{m}}^c \tilde{n}^d 
= 
-\frac{M}{\left( r - ia\cos(\theta)\right)^3} 
.
\end{equation}
The proportionality factor will be directly computed 
due to a close relation with the intervening Carter's constant 
and will be shown in section \ref{sec:lensing-Kerr}.

\subsection{The null geodesic equation}\label{subsec:Integrability+geod}
It is well known that in Kerr spacetime, the geodesic equation can be expressed
as a system of first order differential equations\cite{Carter:1968rr,Walker:1970un};
due to the existence of four constants of motion.

In Boyer-Lindquist coordinates, the system of equations for the components
$(\dot{t}, \dot{r}, \dot{\theta}, \dot{\phi})
=
(\ell^t, \ell^r, \ell^\theta, \ell^\phi)$
of $\ell^a$, are: 
\begin{align}
\Sigma \dot t 
=& 
\frac{1}{\Delta}\left[ E \Big(
(r^2 + a^2)^2 - \Delta \, a^2 \sin^2(\theta) \Big)
- 2 a M r L_z\right]
, 
\label{eq:ellt-geod}
\\
\Sigma^2 \dot r^2 =& \mathcal{R}
, 
\label{eq:ellr-geod}
\\
\Sigma^2 \dot \theta^2 =& \varTheta 
,
\label{eq:elltheta-geod}
\\
\Sigma \dot \phi =& \frac{1}{\Delta}\left[ 2E a M r + (\Sigma - 2Mr) \frac{L_z}{\sin^2(\theta)}\right]
;
\label{eq:ellphi-geod}
\end{align}
where 
the functions $\mathcal{R}$ and $\varTheta$ are defined as
\begin{align}
\mathcal{R} =& \Big( E\left(r^2 + a^2 \right) - a L_z \Big)^2 - K \Delta 
,
\label{eq:R+geod+func}
\\
\varTheta =& K - \left( \frac{L_z}{\sin(\theta)} - a E \sin(\theta) \right)^2 ,
\end{align}
and $E=g(\xi_t,\ell)$ and $L_z=-g(\xi_\phi,\ell)$ are the conserved energy and the angular momentum component
of the photons,
with respect to the Killing vectors $\xi_t \equiv \frac{\partial}{\partial t}$ and
$\xi_\phi \equiv \frac{\partial}{\partial \phi}$,
while $K$ denotes Carter's constant which can be expressed as:
\begin{equation}\label{eq:carterk}
K =
2 \Sigma \ell^a \ell^b \tilde{\ell}_a \tilde{n}_b 
;
\end{equation}
where $\tilde{\ell}^a$ and $\tilde{n}^b$ are the principal null vectors presented above.

It is also useful to have at hand the expression of the geodesic co-vector $\ell_a $, namely:
\begin{equation}
\begin{split}\label{eq:dele-general}
\ell_a =& 
\,E dt_a 
- 
\frac{\pm 	\sqrt{\mathcal{R} }
}{\Delta} dr_a 
- \big( \pm \sqrt{ \Theta }   \big)  d\theta_a 
- L_z d\phi_a
.
\end{split}
\end{equation}

Let us note that due to our choice of  parameterization \eqref{eq:normalization+lambda},
and that at the observer position we are interested in `lensed' light rays,
one has
\begin{equation}
\left. \dot r \right|_o = \left. \frac{\sqrt{\mathcal{R}}}{\Sigma} \right|_o
,
\end{equation}
and
\begin{equation}
\left. \dot \theta \right|_o = \left. \frac{\pm\sqrt{\varTheta}}{\Sigma} \right|_o
;
\end{equation}
where we use the subindex $_o$ to indicate at the observer position.

We think of the observer to be at 
a spacetime event $p_o$, with four velocity $v^a$, at the moment in which
an astronomical picture is taken in some direction.
Each null geodesic in the bundle reaching his/her camera is completely 
determined by the constants $E$, $L_z$ and $K$; mentioned above.
In other, words, these constants are related to the position
and to the angular coordinates in the 
sky used by the observer.
These relations are studied next.

\section{The constants of motion in terms of observed angles}\label{sec:Integr+Const}

We intend our work to be of practical use in astrophysical studies,
and therefore we will provide relations between the observed angles
and observer position with
the quantities appearing in the formalism.
Then, in this section we establish the relations between the 
constants of motion of the null geodesics with the observed angles
with respect to a center in the sphere of directions.
We define it, recurring to
the center of mass null geodesic\cite{Arganiaraz19a} passing through the observer
position and directed to the tilted black hole with angular momentum $J=aM$.
This in turn provides a relation between the natural frame
defined by the observer with that given by the Kerr geometry.

We choose the observer to have a
stationary motion, which assigns to the lens a position $(x_l, y_l,z_l) = (0,d_l,0)$ 
in a local Cartesian system where the $y$ direction is associated to the
spacelike projection of the center of mass null geodesic.
In this frame, one is led to define a notion of tilted angle for the black-hole since
for an arbitrary position of the observer, the rotation axis will appear 
tilted an angle $\iota$, towards the observer,
with respect to the local $z$-direction in the frame of the observer.
Since the spacetime is not flat, the observer is only allowed
to use this notation in its tangent spacetime;
however this is a practical concept suitable for all astrophysical observations.
Below we will give a precise geometrical meaning to these concepts just mentioned.

\subsection{Relation of the observer frame with Boyer-Lindquist frame}\label{subsec:Frames}

In the following we prescribe our particular choice of tetrad 
$(\mathsf{T}^a, \mathsf{X}^a, \mathsf{Y}^a, \mathsf{Z}^a)$ for an observer 
at the event $(t_o, r_o, \theta_o, \phi_o)$, given in Boyer-Lindquist coordinates.
Due to axial-symmetry, and the periodic nature of coordinate $\phi$,
we are free to choose $\phi_o$; in our setting we will take 
$\phi_0 = -\frac{\pi}{2}$.	
We will choose this tetrad in such way that the direction $\mathsf{Y}^a$ points out to
the `center'of the black-hole. 
Our specific choice of tetrad is described  next.

First of all, we define an orthonormal tetrad of Boyer-Lindquist system 
$(\hat{\mathsf{T}}^a, \hat{\mathsf{R}}^a, \hat{\mathsf{\Theta}}^a, \hat{\mathsf{\Phi}}^a)$, 
given by:
\begin{align}
\hat{\mathsf{T}}^a =& 
\frac{1}{\sqrt{1 - \Phi}} \partial_t^a 
,
\\
\hat{\mathsf{R}}^a =& \sqrt{\frac{\Delta}{\Sigma}} \partial_{r}^a
,
\\
\hat{\mathsf{\Theta}}^a =& \sqrt{\frac{1}{\Sigma}} \partial_{\theta}^a
,
\\
\hat{\mathsf{\Phi}}^a =&
\frac{\sqrt{1 - \varPhi}}{\sin(\theta) \sqrt{\Delta} }
\bigg( \partial_{\phi}^a 
- \frac{\varPhi \,  a  \sin^2(\theta)}{1 - \varPhi} \partial_t^a \bigg) 
.
\end{align}

For future reference let us note from the last equation one deduces
\begin{equation}
\partial_{\phi}^a
=
\frac{\varPhi \,  a  \sin^2(\theta)}{\sqrt{1 - \varPhi}} \hat{\mathsf{T}}^a  
+
\frac{\sin(\theta) \sqrt{\Delta} }{\sqrt{1 - \varPhi}} \hat{\mathsf{\Phi}}^a 
.
\end{equation}

Lowering the indices one has
\begin{align}
\hat{\mathsf{T}}_a =& \sqrt{1 - \varPhi} dt_a 
+ \frac{\varPhi a \sin(\theta)^2}{\sqrt{1 - \varPhi}}  d\phi_a
,
\\
\hat{\mathsf{R}}_a =& -\sqrt{\frac{\Sigma}{\Delta}} dr_a 
,
\\
\hat{\mathsf{\Theta}}_a =& -\sqrt{\Sigma} d\theta_a
,
\\
\hat{\mathsf{\Phi}}_a =& 
- \frac{\sqrt{\Delta}}{\sqrt{1 - \varPhi}}\sin(\theta)d\phi_a
.
\end{align}

Our first element $\mathsf{T}^a$ for the observer frame is
\begin{equation}
\mathsf{T}^a = \hat{\mathsf{T}}^a
= 
\frac{1}{\sqrt{1 - \Phi_o}} \partial_t^a
.
\end{equation}

The spacelike direction $\mathsf{Y}^a$, as we said before, will point to the `center' of the 
black-hole. 
But, since the black hole has angular momentum, it is not immediate what should be
considered the `center' direction.
Furthermore, since in general, the observer will see the black-hole 
as tilted an angle $\iota$, the shape of the black hole as seeing
by the observer, will not have a symmetric form.
The correct way to choose this `center' 
is to identify the direction of $\mathsf{Y}^a$ 
with the spacelike direction of
the unique null geodesic
that belongs to the center of mass(\emph{cm}) null congruence, as described in the work 
\cite{Arganiaraz19a}.
This congruence is represented by the directions 
\eqref{eq:ellt-geod} -  \eqref{eq:ellphi-geod} where $L_z=0$ and the function 
$K_\text{cm}(r_o,\theta_o)$
has been numerically calculated in \cite{Arganiaraz19a}.
We denote the function $K$ in \cite{Arganiaraz19a} as $K_\text{cm}$,
to take into account the fact that there the parameter $E$ was chosen to
have the unit value.

In those studies, it comes naturally the suggestion\cite{Moreschi18a} to define $\tilde{k}$
from the relation
\begin{equation}\label{eq:kchica2}
K_\text{cm}(r,\theta) = \tilde K - \tilde k(r,\theta)^2 ;
\end{equation}
where $\tilde K$ is the function $K_\text{cm}$ in the limit for the mass
$M\rightarrow 0$, and is:
\begin{equation}\label{eq:ktilde}
\tilde K(r,\theta)
=
\frac{(r^2 + a^2)\, a^2 \sin(\theta)^2 }{r^2 + a^2 \sin(\theta)^2}
.
\end{equation}
See appendix \ref{ap:Center+mass+Kfunction} for further details.
Suppose one would like to apply the above equations to the case
of the supermassive black hole in the center of our galaxy.
Its mass is about $M =8.2 \times 10^{36}$kg; 
which in units of length is 
$\frac{G}{c^2} M = \frac{6.674 \times 10^{-11}}{(3 \times 10^8)^2}8.2 
\times 10^{36}\text{m}=6.08 \times 10^{9}$m.
While its distance is about $d = 26000\text{ly}=2.46\times 10^{20}$m;
so that approximately $d  = 4.04 \times 10^{10} M$.
For other black holes of mass $m$ in the galaxy one would have masses $m<<M$.
While for other supermassive black holes in the cosmos
at distance $D$, one would have $D>>d$ .
We conclude then, that for black holes of astrophysical interest
of mass $M$ and angular momentum $a$, they will normally be
at distances satisfying $d > 10^{10} M$.
Then, in \eqref{eq:ktilde} we can make the approximation
\begin{equation}
\begin{split}
\frac{(r_o^2 + a^2) }{r_o^2 + a^2 \sin(\theta_o)^2}
=&
\frac{(1 + \frac{a^2}{r_o^2} ) }{1 + \frac{a^2 \sin(\theta_o)^2}{r_o^2} }
\\
\cong &
\left(1 + \frac{a^2}{r_o^2} \right) \left(1 - \frac{a^2 \sin(\theta_o)^2}{r_o^2}\right) \\
\cong&
1 + \frac{a^2 \cos(\theta_o)^2}{r_o^2} + \mathscr{O}\left(\frac{a^4}{r_o^4}\right)
\\
\cong & 
1 + \frac{a^2 \cos(\theta_o)^2}{M^2} 10^{-20}
;
\end{split}
\end{equation}
that is, we can just use the unit value for all practical
astrophysical applications.
So that we conclude that in astrophysical applications
one can use
$\tilde K(r,\theta)\cong a^2 \sin(\theta)^2$;
for the value of $\tilde K$ at the observer position.

By inspection in the numerical calculations of \cite{Moreschi18a},
we see that the maximum numerical value of $\tilde k$
at $\xi_1 = 1.220703E-03$ is
$\tilde k_\text{max}(\xi_1) = 7.254817E-09$; 
where $\xi=1/r$ (For a unit mass black hole.).
Recalling the astrophysical situation mentioned above of an
observer at an approximate radial position of $d = 10^{10} m$;
it is deduced that, assuming a linear interpolation, at the observer
position $\xi(d)= 10^{-10}$ one would have
$\tilde k_\text{max}(\xi(d)) 
= \tilde k_\text{max}(\xi_1) \xi(d)/\xi_1 = 5.94314E-22$.

Taking into account the previous discussion on $\tilde K$,
for large values of the radial coordinate,
we conclude that in astrophysical applications
one can use
\begin{equation}
K_\text{cm}(r=d,\theta)= a^2 \sin(\theta)^2+\delta
;
\end{equation}
at the observer position,
where $\delta/m^2 = \mathscr{O}(10^{-20})$.
In terms of our setting, if we observe \eqref{eq:elltheta-geod} we see that $K_{o_\text{cm}} = E^2 K_\text{cm}$;
so that we take the working value of $K_{o_\text{cm}}$, for the reference center of mass
null geodesic, at the observer position
to be
\begin{equation}
K_{o_\text{cm}} = E^2 a^2 \sin(\theta_o)^2
;
\end{equation}
where $\theta_o$ is the value of the coordinate $\theta$ at the observer
position.

Regarding the angular coordinate of the observer, if we denote with
$\tilde \theta$ the standard azimuthal angle in spherical
coordinates, in the limit for the black hole mass $M\rightarrow 0$;
then, we also deduce that at this location, one has
\begin{equation}
\tilde \theta_o = \theta_o + \epsilon_d ;
\end{equation}
with $\epsilon_d = \mathscr{O}(10^{-10})$.
This is closely connected to the inclination
of the axis of 
the black hole as seen by the observer.
Since the spacetime is curved and the observer is supposed to be
at a finite distance, one has to agree in the notion of the 
\emph{tilted angle} of the black hole. 
In particular, it should not be tied to a local
coordinate; but we have just seen that
the Boyer-Lindquist coordinate $\theta_o$ and the flat background
coordinate $\tilde{\theta}_o$, 
agree with high precision
for astrophysical observations; for this reason,
if the observer is at the angle $\theta_o$,
we employ
\begin{equation}\label{eq:iota}
\iota = \frac{\pi}{2} - \theta_o 
;
\end{equation}
as an exact expression for the tilted angle 
in astrophysical use.

Let us note that from equation (\ref{eq:elltheta-geod}), one has for all practical
purposes, $\dot \theta = 0$ at the observer position for the \emph{cm} geodesic.

Therefore, equations (\ref{eq:ellr-geod}) and (\ref{eq:ellphi-geod}) led us to
take
\begin{equation}
\begin{split}
\mathsf{Y}^a \equiv& 
- \frac{\sqrt{\Delta_o}}{N_{\mathsf{y}}}
\left[
\sqrt{\frac{\mathcal{R}_\text{cm} }{\Sigma_o \Delta_o}}\hat{\mathsf{R}}^a +
\frac{\varPhi_o a \sin(\theta_o)}{\sqrt{\Delta_o}\sqrt{1 - \varPhi_o}}
\hat{\mathsf{\Phi}}^a
\right] 
\\
=&
- \frac{\sqrt{\Delta_o}}{N_{\mathsf{y}}}
\left[
\frac{\sqrt{\mathcal{R}_\text{cm} }}{\Sigma_o } \partial_{r}^a  
+
\frac{a \varPhi_o }{\Delta_o}
\partial_{\phi}^a 
- \frac{a^2 \varPhi_o^2 \,  \sin^2(\theta_o)}{\Delta_o (1 - \varPhi_o)} \partial_t^a
\right] 
;
\end{split}
\end{equation}
where $\mathcal{R}_\text{cm}$ is given by 
\begin{equation}
\begin{split}
\mathcal{R}_\text{cm} =& E^2\left(r_o^2 + a^2 \right)^2 - K_{o_\text{cm}} \Delta_o 
;
\end{split}
\end{equation}  
and $N_{\mathsf{y}}$ is a normalization factor given by
\begin{equation}
\begin{split}
N_{\mathsf{y}} =& \sqrt{\frac{\mathcal{R}_\text{cm} }{\Sigma_o} + 
	\frac{\varPhi_o^2 a^2 \sin(\theta_o)^2}{1 - \varPhi_o}
} 
.
\end{split}
\end{equation}
We have employed here $\mathcal{R}_\text{cm}$ instead of $\mathcal{R}_o$ to 
emphasize that we are considering the function $\mathcal{R}$ in \eqref{eq:R+geod+func} 
at the observer position but taking $L_z = 0$ and $K = K_{o_\text{cm}}$. We make this note at this point since
in the next section we will use $\mathcal{R}_o$ to mean a general position
$(r_o, \theta_o)$, with $L_z \neq 0$ and general $K$ for null geodesics reaching the observer.

We choose at the observer position either
\begin{equation}
\begin{split}
\mathsf{X}^a \equiv& 
\frac{\sqrt{\Delta_o}}{N_{\mathsf{y}}}
\left[
\sqrt{\frac{\mathcal{R}_\text{cm} }{\Sigma_o \Delta_o}}\hat{\mathsf{\Phi}}^a 
-
\frac{\varPhi_o a \sin(\theta_o)}{\sqrt{\Delta_o}\sqrt{1 - \varPhi_o}}
\hat{\mathsf{R}}^a
\right] \\
=&
\frac{\sqrt{\Delta_o}}{N_{\mathsf{y}}}
\bigg[
- \frac{\varPhi_o \,  a  \sin(\theta_o)}{ \Delta_o \sqrt{1 - \varPhi_o}} 
\sqrt{\frac{\mathcal{R}_\text{cm} }{\Sigma_o}}
\partial_t^a
\\
& \qquad \; \, +
\sqrt{\frac{\mathcal{R}_\text{cm} }{\Sigma_o}}
\frac{\sqrt{1 - \varPhi_o}}{\sin(\theta_o) \Delta_o }
\partial_{\phi}^a 
-
\frac{\varPhi_o a \sin(\theta_o)}{\sqrt{\Sigma_o}\sqrt{1 - \varPhi_o}}
\partial_{r}^a
\bigg]
;
\end{split}
\end{equation}
or
\begin{equation}
\mathsf{Z}^a \equiv - \hat{\mathsf{\Theta}}^a
=
- \frac{1}{\sqrt{\Sigma_o}} \partial_{\theta}^a 
;
\end{equation}
where the base vectors of the Kerr geometry are also assumed to
be evaluated at the observer position.
It can be seen that choosing one of these two
base vectors the other is completely determined
by the conditions of orthonormality and orientation.

For further calculation we also include the lower index expressions of the 
observer frame given by
\begin{align}
\mathsf{T}_a =& \sqrt{1 - \varPhi_o} dt_a 
+ \frac{\varPhi_o a \sin(\theta_o)^2}{\sqrt{1 - \varPhi_o}} d\phi_a
,
\\
\mathsf{X}_a =& 
\frac{\sqrt{\Delta_o}}{N_{\mathsf{y}}}\left[ 
\frac{\varPhi_o a \sin(\theta_o)}{\sqrt{\Delta_o} \sqrt{1 - \varPhi_o}} 
\sqrt{\frac{\Sigma_o}{\Delta_o}}dr_a - 
\sqrt{\frac{\mathcal{R}_\text{cm} }{\Sigma_o}}
\frac{\sin(\theta_o)}{\sqrt{1 - \varPhi_o}}d\phi_a
\right]
,
\\
\mathsf{Y}_a =& 
\frac{\sqrt{\Delta_o}}{N_{\mathsf{y}}}\left[
\frac{\sqrt{\mathcal{R}_\text{cm}} }{\Delta_o}
dr_a +
\frac{\varPhi_o a \sin(\theta_o)^2}{1 - \varPhi_o} 
d\phi_a
\right]
,
\\
\mathsf{Z}_a =& 
\sqrt{\Sigma_o}d\theta_a
.
\end{align}

Let us realize that, with this choice of tetrad the plane orthogonal to the line of sight 
is expanded by the directions $\mathsf{X}^a$ and $\mathsf{Z}^a$.
In particular, it will be natural to use the angles $(\alpha_{x}, \delta_{z})$
in the observed images, which represent
the departure from a preferred direction, namely the center of mass one.
Assuming a flat geometry, the observers will assign the cartesian coordinates:
\begin{align}
x_s \equiv& \lambda_s\sin\left(\frac{\pi}{2} - \delta_{z} \right) 
\cos\left( \frac{\pi}{2} - \alpha_{x} \right) ,\\
z_s \equiv& \lambda_s \cos\left(\frac{\pi}{2} - \delta_{z} \right).
\end{align}
to a point like object situated at an affine distance $\lambda_s$.
We discuss in more detail the observed angles in the next subsection.

\subsection{Angular coordinates of the null geodesics as seen in the observer frame}
At the place of the observer, let us consider an orthonormal frame, namely
$(\mathsf{T}^a, \mathsf{X}^a, \mathsf{Y}^a, \mathsf{Z}^a)$
as described above.
The observer is assumed to be stationary with respect
to observers at infinity; so that its own velocity is
\begin{equation}
v^a = \mathsf{T}^a 
= \frac{1}{\sqrt{g_{tt}}} \partial_t^a 
= \frac{1}{\sqrt{1 - \varPhi_o}} \partial_t^a 
.
\end{equation}

Let us note that from \eqref{eq:normalization+lambda} one has
\begin{equation}
1 = 
\ell_a \mathsf{T}^a 
= \frac{E}{\sqrt{1 - \varPhi_o}}
;
\end{equation}
which indicates that
\begin{equation}
E = \sqrt{1 - \varPhi_o}
.
\end{equation}
Note that, in order to simplify the reading, 
we have not used a subindex for the constant $E$ since it only depends
on the observer position and therefore it is the same for all geodesics in the
past null cone of the observer.

We define 
the \emph{angular coordinates in the sphere of directions of the observer}, 
namely $(\alpha_x, \delta_{z})$, in the following way.
For a null geodesic $\ell^a$ reaching the observer we set:
\begin{align}
\sin\left(\frac{\pi}{2} - \delta_{z} \right) \cos\left( \frac{\pi}{2} 
- \alpha_x \right) \equiv& 
~\ell^a \mathsf{X}_a 
, \label{eq:ell+hatX+prima}\\
\sin\left(\frac{\pi}{2} - \delta_{z} \right) \sin\left( \frac{\pi}{2} 
- \alpha_x \right) \equiv& 
~\ell^a \mathsf{Y}_a 
, \label{eq:ell+hatY+prima}\\
\cos\left(\frac{\pi}{2} - \delta_{z} \right) \equiv& 
~\ell^a \mathsf{Z}_a 
. \label{eq:ell+hatZ+prima}
\end{align} 
In particular, for $(\alpha_x, \delta_{z}) = (0,0)$, the observer
will be pointing the telescope to a direction that we take as that of the
center of mass null geodesic, passing through the observer position;
explained above.

Therefore, a null geodesic vector $\ell^a$ can be written in the tangent space 
of the observer as
\begin{equation}\label{eq:ell+observer}
\begin{split}
\ell^a =&  \mathsf{T}^a 
{-}
\sin\left(\frac{\pi}{2} - \delta_{z} \right) \cos\left( \frac{\pi}{2} 
- \alpha_{x} \right) \mathsf{X}^a 
\\
&
{-}
\sin\left(\frac{\pi}{2} - \delta_{z} \right) \sin\left( \frac{\pi}{2} 
- \alpha_{x} \right) \mathsf{Y}^a 
{-}
\cos\left(\frac{\pi}{2} - \delta_{z} \right) \mathsf{Z}^a.
\end{split}
\end{equation}
\begin{figure}
	\centering
	\includegraphics[clip,width=0.35\textwidth]{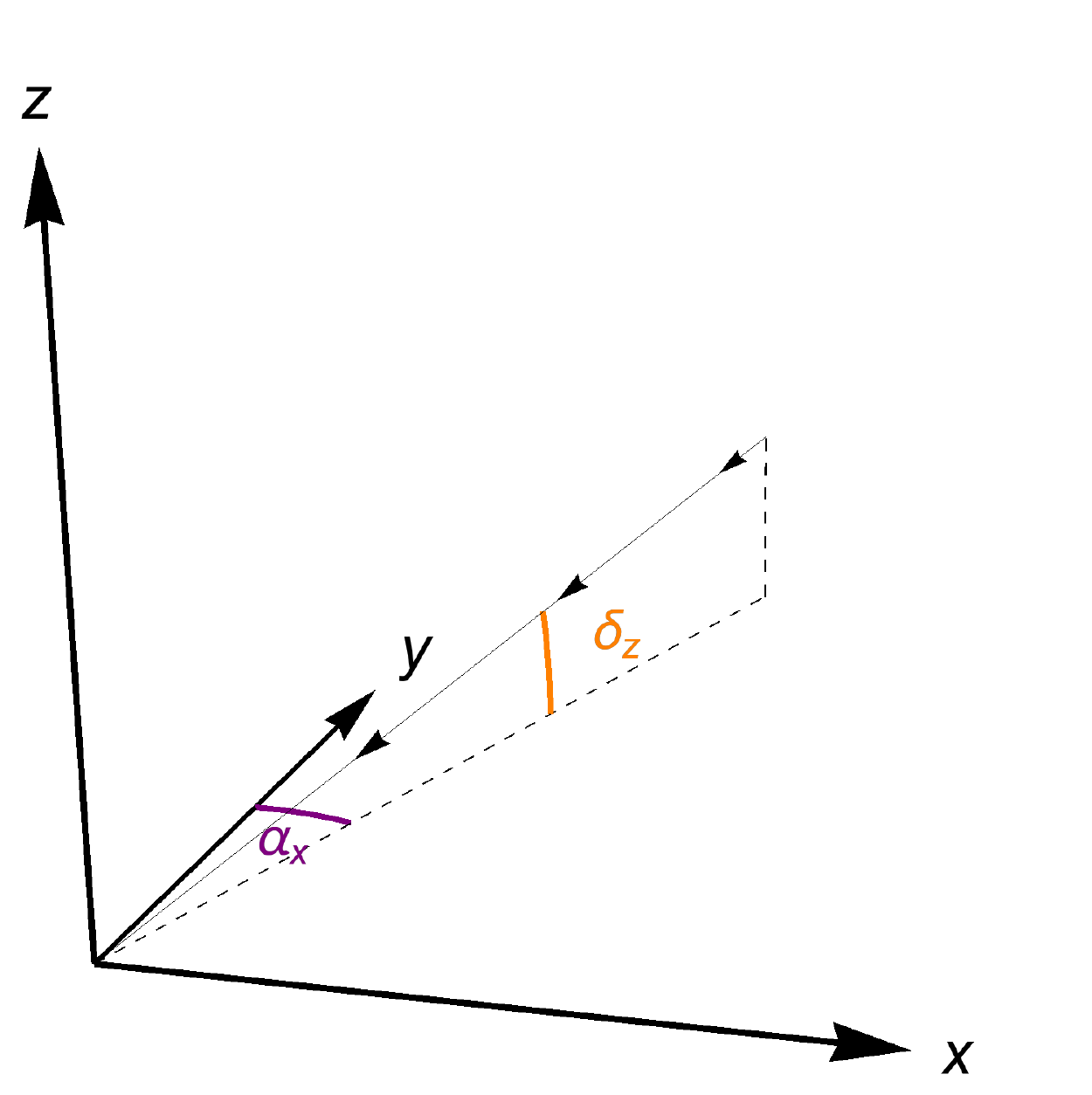}
	\caption{The figure shows the reference system of the observer and the angular
		coordinates employed to describe the directions of incoming photons.
	}\label{fig:Sky_Directions_Lensing}
\end{figure}

Employing the reference frame 
$(\mathsf{T}^a, \mathsf{X}^a, \mathsf{Y}^a, \mathsf{Z}^a)$
of the observer and the angular coordinates $(\alpha_{x}, \delta_{z})$
we can also consider a null tetrad of the form described in section \ref{subsec:Geod+Dev} 
which is adapted to the null geodesic $\ell^a$ itself.
In fact, we take the null vectors $(n^a, m^a, \bar{m}^a)$ as follows:
\begin{equation}\label{eq:n+observer}
\begin{split}
n^a =&  \frac{1}{2} \mathsf{T}^a 
{+}
\frac{1}{2}\sin\left(\frac{\pi}{2} - \delta_{z} \right) 
\cos \left( \frac{\pi}{2} - \alpha_{x} \right) \mathsf{X}^a \\
& 
{+}
\frac{1}{2}\sin\left(\frac{\pi}{2} - \delta_{z} \right) 
\sin\left( \frac{\pi}{2} - \alpha_{x} \right) \mathsf{Y}^a 
{+}
\frac{1}{2}\cos\left(\frac{\pi}{2} - \delta_{z} \right) \mathsf{Z}^a ,
\end{split}
\end{equation}
\begin{equation}\label{eq:m+observer}
\begin{split}
m^a =& \frac{1}{\sqrt{2}}
\left[
{-} i \cos\left(\frac{\pi}{2} - \delta_{z} \right) 
\cos\left( \frac{\pi}{2} - \alpha_{x} \right)
+ \sin\left( \frac{\pi}{2} - \alpha_{x} \right) \right] \mathsf{X}^a \\
&-  \frac{1}{\sqrt{2}}
\left[
 i \cos\left(\frac{\pi}{2} - \delta_{z} \right) 
\sin\left( \frac{\pi}{2} - \alpha_{x} \right) 
+ \cos\left( \frac{\pi}{2} - \alpha_{x} \right) \right] \mathsf{Y}^a 
\\
& + \frac{i}{\sqrt{2}}\sin\left(\frac{\pi}{2} - \delta_{z} \right) \mathsf{Z}^a,
\end{split}
\end{equation}
\begin{equation}
\begin{split}
\bar{m}^a =& \frac{1}{\sqrt{2}}
\left[
i \cos\left(\frac{\pi}{2} - \delta_{z} \right) 
\cos\left( \frac{\pi}{2} - \alpha_{x} \right)
+  \sin\left( \frac{\pi}{2} - \alpha_{x} \right) \right] \mathsf{X}^a 
\\
& + \frac{1}{\sqrt{2}}
\left[
i \cos\left(\frac{\pi}{2} - \delta_{z} \right) 
\sin\left( \frac{\pi}{2} - \alpha_{x} \right)
-  \cos\left( \frac{\pi}{2} - \alpha_{x} \right) \right] \mathsf{Y}^a 
\\
& - \frac{i}{\sqrt{2}}\sin\left(\frac{\pi}{2} - \delta_{z} \right) \mathsf{Z}^a.
\end{split}
\end{equation}

This choice has the property that when evaluated at the center of mass null
geodesic, the complex null vector $m^a$ is orthogonal to $\mathsf{Y}^a$,
and also it agrees with the conventions as given in \cite{Boero:2016nrd}.

A null tetrad along the geodesic characterized by 
$(\alpha_{x}, \delta_{z})$,
coming from the source, is obtained by parallel transporting 
the four vectors, defined above at the observer position.
Let us note that, in fact, this is the null tetrad that we use in order to compute the 
curvature scalars which appear in the geodesic deviation equation in the exact lensing
calculations.

\subsection{Relation between the constants of motion and the observed angles}

Now we are ready to relate the angular variables $(\alpha_{x}, \delta_{z})$ 
for an arbitrary null `lensed' geodesic reaching the observer with the
constants of motion.

Using expressions (\ref{eq:ell+hatX+prima} - \ref{eq:ell+hatZ+prima}) we obtain
\begin{equation}\label{eq:cos-deltaz}
\cos\left(\frac{\pi}{2} - \delta_{z} \right) = 
\sqrt{\Sigma}_o\ell^\theta
=  \frac{\pm \sqrt{\varTheta}}{\sqrt{\Sigma_o} },
\end{equation}
and
\begin{equation}\label{eq:cos-alfax}
\begin{split}
\cos\left( \frac{\pi}{2} - \alpha_{x} \right)
=& 
\frac{\ell^a \mathsf{X}_a}{\sqrt{1 - \frac{\varTheta}{ \Sigma_o}}} 
= 
\frac{\left(1 - \frac{\varTheta}{ \Sigma_o}\right)^{-1/2} \sin(\theta_o) 
	\sqrt{\Delta_o}}{N_\mathsf{y} \sqrt{1 - \varPhi_o}}
\\
&
\left[ 
\frac{\varPhi_o a \sqrt{\Sigma_o}}{\Delta_o} \ell^r 
-  \frac{\sqrt{\mathcal{R}_\text{cm}} }{\sqrt{\Sigma_o}} \ell^\phi \right]
\\
=& 
\frac{\left(1 - \frac{\varTheta}{\Sigma_o}\right)^{-1/2} \sin(\theta_o)}{N_\mathsf{y} \sqrt{\Delta_o} \sqrt{1 - \varPhi_o}}
\left[ 
\frac{\varPhi_o a }{ \sqrt{\Sigma_o}} \left( \sqrt{\mathcal{R}(K)}\right) 
\right.
\\
& \left. -  \frac{\sqrt{\mathcal{R}_\text{cm}} }{\sqrt{\Sigma_o}}  
\left( Ea \varPhi_o + \left(1 - \varPhi_o \right)\frac{L_z}{\sin(\theta_o)^2}\right)
\right]
.
\end{split}
\end{equation}

These are the exact relations linking the observed angles 
$(\alpha_{x}, \delta_{z})$ to the constant of 
motion $K \equiv K(\alpha_{x}, \delta_{z})$, $L_z \equiv L_z(\alpha_{x}, \delta_{z})$
and $E$.

For practical purposes it is worthwhile to know the behavior of equations 
(\ref{eq:cos-deltaz}-\ref{eq:cos-alfax}) when 
the position of the observer is such that $M/r_0 \ll 1$ and also  $a/r_0 \ll 1$.
In addition one will be dealing with $|\alpha_{x}| \ll 1$ and 
$|\delta_{z}| \ll 1$ and therefore:
\begin{align}
\alpha_{x}
=&
- 
\frac{L_z}{r_o \sin(\theta_o)}
+ \mathscr{O}(\frac{\Phi_o L_z}{r_o} )
, 
\\
\begin{split}
\delta_{z} 
=&
\frac{(\pm)}{r_o} 
\left[K - \left( \frac{L_z}{\sin(\theta_o)} - aE\sin(\theta_o) \right)^2\right]^{1/2}
+ \mathscr{O}(\frac{a^2 \sqrt{\Theta}}{r_o^3} )
\\
=&
\frac{(\pm)}{r_o} 
\left[K 
- \left( \sqrt{K_{o_\text{cm}}} - \frac{L_z}{\sin(\theta_o)}  \right)^2
\right]^{1/2}
+ \mathscr{O}(\frac{a^2 \sqrt{\Theta}}{r_o^3} )
.
\end{split}
\end{align} 
Since the argument in this square root can not be negative,
we can define a quantity $d^2$
so that $K$ is expressed as
\begin{equation}
K = K_{o_\text{cm}}  
+
2\alpha_x r_o \sqrt{K_{o_\text{cm}}} + d^2 r_o^2 
;
\end{equation}
from which we obtain
\begin{equation}
\delta_{z} 
=
(\pm) \sqrt{d^2 - \alpha_x^2  }
+ \mathscr{O}(\frac{a^2 \sqrt{\Theta}}{r_o^3} )
,
\end{equation}
and we also see that one must have 
$d^2 \geqslant \alpha_x^2$;
due to the triangular relation, 
since 
$d^2 \approx \alpha_x^2 + \delta_{z}^2$ 
represents the angular distance to the reference center of mass direction.

These expressions for small angles in the asymptotic region are similar to
those found by Chandrasekar in its discussion of geodesic motion in \cite{Chandrasekhar:1985kt}.

Expressing the constant of motion in terms of the observed angles, one has
\begin{equation}
L_z =
- 
\alpha_x r_o \sin(\theta_o)
+ \mathscr{O}(\Phi_o L_z )
,
\end{equation}
and
\begin{equation}
K = K_{o_\text{cm}} 
+
2\alpha_x r_o \sqrt{K_{o_\text{cm}}} 
+ (\alpha_x^2 + \delta_z^2 ) r_o^2 
+ \mathscr{O}( \frac{a^2 \sqrt{\Theta} \delta_z}{r_o} )
;
\end{equation}
where it could be remarked that $\delta_z r_o / M > 1$.

\section{Calculation of the lensing curvature component $\Psi_{0}$ in Kerr spacetime}\label{sec:lensing-Kerr}

Let us recall that since Kerr geometry is a vacuum solution, 
the curvature matrix appearing in equation \eqref{eq:matrix+Q} is 
completely determined by $\Psi_0$. 

Here we show how to obtain an exact expression for the Weyl scalar $\Psi_0$ 
without the need to know explicitly the null tetrad adapted to the geodesic 
motion of photons (as described in section \ref{sec:Integr+Const})
along the whole trajectory, but only at the position of the observer,
and so
avoiding the lengthy computation of contractions in equation \eqref{eq:Psi0+def}.
The simplicity of the result constitutes a curious although useful improvement 
on the literature of gravitational lensing on Kerr spacetime; in particular 
to pioneering work such as \cite{Pineault1977ApJ,Pineault1977ApJ_II}.

\subsection{Relations between the principal null tetrad and the null tetrad adapted to the 
	photons}

Let us start by recalling that $\Psi_0$ must be proportional to $\tilde{\Psi}_2$
due to the fact that Kerr is algebraically type-D.
The proportionality function depends on a Lorentz transformation between the two
null tetrad involved in the definitions of both curvature scalars.

A \emph{restricted Lorentz transformation} (see appendix \ref{ap:hom-Lorentz-trans}) 
between the principal null tetrad $(\tilde{\ell}^a, \tilde{m}^a, \bar{\tilde{m}}^a ,\tilde{n}^a)$ 
and $(\ell^a, m^a , \bar{m}^a, n^a)$ has the form:
\begin{align}
\ell^a =& 
Z\tilde{\ell}^a + Z\varLambda \bar{\tilde{m}}^a + Z\bar{\varLambda}\tilde{m}^a + 
Z\varLambda \bar{\varLambda} \tilde{n}^a , \label{eq:ell-Lorentz} 
\\
\begin{split}
m^a =& 
\varGamma Z \tilde{\ell}^a + \varGamma Z \varLambda \bar{\tilde{m}}^a 
+ 
\left( \varGamma Z \bar{\varLambda} + e^{is} \right)\tilde{m}^a 
\\
& + 
\varLambda \left( \varGamma Z  \bar{\varLambda} +  e^{is} \right) \tilde{n}^a ,\label{eq:m-Lorentz}
\end{split}
\\
\begin{split}
n^a =& 
\varGamma \bar{\varGamma} Z \tilde{\ell}^a 
+ 
\left( \bar{\varGamma} e^{is} + \varGamma \bar{\varGamma}
Z \bar{\varLambda} \right) \tilde{m}^a 
\\
& +
\left( \varGamma e^{-is} + \bar{\varGamma} \varGamma
Z  \varLambda \right) \bar{\tilde{m}}^a 
\\
&+ 
\left( \frac{1}{Z} + \varGamma \bar{\varLambda}e^{-is} + \bar{\varGamma} \varLambda e^{is} + 
\varGamma \bar{\varGamma} Z 
\varLambda \bar{\varLambda} \right)\tilde{n}^a; \label{eq:n-Lorentz}
\end{split}
\end{align}
being $Z$ and $s$ real functions and $\varGamma$ and $\varLambda$ complex ones.

Restricted Lorentz transformations preserve the choice of orientation on the spacetime; 
i.e. the sign of the volume element $\epsilon_{abcd}$; and therefore we have taken care 
of this fact previously in our choice for the principal null tetrad.
Let us realize that for distant observers the null vector $\ell^a$ and the complex null 
vector $m^a$ behave as
\begin{align}
\ell^a &\sim E \partial_t + E \partial_r, \label{eq:ell+asyntotico}\\
m^a &\sim -i\left( \frac{1}{\sqrt{2}r}\partial_\theta + \frac{i}{\sqrt{2}r \sin(\theta)}\partial_\phi \right);
\label{eq:m+asyntotico}
\end{align} 
with $E= 1+\mathscr{O}(\frac{M}{r_o})$,
while the null vectors of the principal null tetrad $\tilde{\ell}^a$ and $\tilde{m}^a$ 
behaves as
\begin{align}
\tilde{\ell}^a &\sim \partial_t + \partial_r , 
\\
\tilde{m}^a &\sim  \frac{1}{\sqrt{2}r} \partial_\theta 
+
\frac{i}{\sqrt{2}r\sin(\theta)}\partial_\phi.
\end{align}
That is, in the asymptotic region one has that:
$\ell^a \propto \tilde{\ell}^a$ and $m^a \propto - i\tilde{m}^a$ and so
the relations (\ref{eq:ell-Lorentz})-(\ref{eq:n-Lorentz}) can be applied in a consistent way.

The transformation (\ref{eq:ell-Lorentz})-(\ref{eq:n-Lorentz}) induces a change in 
the curvature scalars; 
in particular, for $\Psi_0$ we obtain the following:
\begin{equation}
\Psi_0 = 6 (Z \varLambda )^2 \tilde{\Psi}_2 \, e^{2is}.
\end{equation}

We observe that an exact knowledge of $\Psi_0$ just requires to compute the product $Z\varLambda e^{is}$.
The following result\cite{Chandrasekhar:1985kt, Walker:1970un} 
(equation (12) of section $\S 60$ Chapter 7) which holds for all vacuum
algebraically type-D spacetimes, allow us to do it in a relatively easy way:
\begin{teo}\label{teo+1}
	If $\ell^a$ is an affinely parametrized null geodesic vector and $k^a$ an orthogonal vector 
	to $\ell^a$ which is parallelly propagated along it; then, in a vacuum type-D spacetime the following
	quantity is conserved along the geodesic:
	\begin{equation}\label{eq:theorem-K}
	\mathbb{K}_{(k)} = 2\left[(\ell^a \tilde{\ell}_a )(k^a \tilde{n}_a) - 
	(\ell^a \tilde{m}_a)(k^a \bar{\tilde{m}}_a) \right] \tilde{\Psi}_2^{-1/3}.
	\end{equation}
\end{teo}

Taking $\ell^a$ as the affine null geodesic of interest, then the vectors 
$\ell^a$, $m^a$ and $\bar{m}^a$ of our null tetrad $(\ell^a, m^a , \bar{m}^a, n^a)$ 
adapted to the photon paths, 
satisfy the 
hypothesis of the theorem. 

When we consider the conserved quantity $\mathbb{K}_{(m)}$ associated with 
$k^a = m^a$
we remarkably found that the constant of the theorem has a key expression which
help us to obtain $\Psi_{0}$:
\begin{corollary}\label{cor:1}
	For a restricted Lorentz transformation of the form (\ref{eq:ell-Lorentz} - \ref{eq:n-Lorentz})
	the conserved quantity along null geodesics $\mathbb{K}_{(k)}$ of theorem (\ref{teo+1}) vanishes
	for $k^a = \ell^a$ and $k^a = \bar{m}^a$; this is:
	\begin{align}
	\mathbb{K}_{(\ell)} =& 
	2 \left[(Z \varLambda\bar{\varLambda} )(Z) - (Z\varLambda)(Z \bar{\varLambda}) \right] 
	\tilde{\Psi}_2^{-1/3} 
	= 0,
	\\
	\mathbb{K}_{(\bar{m})} =& 
	2 \left[(Z \varLambda\bar{\varLambda} )(Z \bar{\varGamma}) - (Z\varLambda)(\bar{\varGamma} Z \bar{\varLambda}) 
	\right] \tilde{\Psi}_2^{-1/3} 
	= 0,
	\end{align}
	and has the following value for $k^a = m^a$:
	\begin{equation}\label{eq:theorem}
	\begin{split}
	\mathbb{K}_{(m)} =& 
	2 \left[(Z \varLambda\bar{\varLambda} )(Z \varGamma) - (Z\varLambda)(\varGamma Z \bar{\varLambda} + e^{is}) \right] \tilde{\Psi}_2^{-1/3} 
	\\
	=& 
	-2 Z\varLambda e^{is} \tilde{\Psi}_2^{-1/3};
	\end{split}
	\end{equation}
	and therefore the Weyl curvature scalar $\Psi_0$ is
	\begin{equation}
	\Psi_0 = \frac{3}{2} \mathbb{K}_{(m)}^2\tilde{\Psi}_2^{5/3}.
	\end{equation}
	
\end{corollary}	
It should be emphasized that this corollary is valid for all vacuum spacetimes of type D.

When we apply this to Kerr spacetime then we obtain:
\begin{corollary}\label{cor:2}
In Kerr spacetime the curvature scalar $\Psi_{0}$ appearing in the 
\emph{geodesic deviation equation} \eqref{eq:dev+geoII} has the 
following expression in terms of Boyer-Lindquist coordinates:
\begin{equation}\label{eq:Psi0+theorem}
\Psi_0 = - \frac{3 M^{5/3} \mathbb{K}_{(m)}^2}{2 \big(r - ia \cos(\theta) \big)^5} 
;
\end{equation}	
where $r=r(\lambda)$, $\theta = \theta(\lambda)$ and 
$\mathbb{K}_{(m)}$ is the conserved quantity of Corollary \ref{cor:1}
\end{corollary}

This is a remarkably simple an exact expression for the curvature in terms of a null tetrad 
adapted to the trajectory of the photons on the Kerr spacetime.
In particular, it will become extremely useful in the combined numerical integration of 
the geodesic and geodesic deviation equations.

It is probably worthwhile to realize that the constant $\mathbb{K}_{(m)}$  
has spin-weight 1, since $m^a$ is an element of our adapted null tetrad, which is consistent
with the fact that $\Psi_0$ is a spin-weight 2 quantity.

The constant $\mathbb{K}_{(m)}$ has been employed for the study of electromagnetic 
polarization in Kerr since it is related to a parallel transported pair of complex
null vectors $\left(m^a, \bar{m}^a\right)$ (see for example \cite{Chandrasekhar:1985kt});
here we show that it is also related to the components of the
curvature that describes the optical scalars in weak lens studies.

\subsection{Relation between the constant $\mathbb{K}_{(k)}$ and Carter's constant $K$}

In this subsection we mention a useful connection between the constant 
$\mathbb{K}_{(k)}$ and Carter's constant $K$,
which has been discussed and proved in Chandrasekhar textbook\cite{Chandrasekhar:1985kt} 
only for real vectors.
Here we extend its validity for the complex vector $m^a$, that is
when $\mathbb{K}_{(k)}=\mathbb{K}_{(m)}$.

First, let us recall the result\cite{Chandrasekhar:1985kt}:
\begin{teo}\label{teo:K0}
A null geodesic, with affine tangent vector $\ell^a$, in any vacuum type-D spacetime 
allows the integral of motion
\begin{equation}\label{eq:K0}
\begin{split}
K_0 =& 2\left| \tilde{\Psi}_2\right|^{-2/3} \left(\ell^a \tilde{m}_a \right) 
\left(\ell^a \bar{\tilde{m}}_a\right) 
\\
=& 
2\left| \tilde{\Psi}_2\right|^{-2/3} \left(\ell^a \tilde{\ell}_a \right) 
\left(\ell^a \tilde{n}_a\right) .
\end{split}
\end{equation}
\end{teo}

The relation between the constant $K_0$ of this theorem and the Carter constant $K$ 
which is present in the equations (\ref{eq:ellt-geod} - \ref{eq:ellphi-geod}),
and given by \eqref{eq:carterk},
is simply
a rescaling by the mass of the spacetime appearing trough $\tilde{\Psi}_2$, that is
\begin{equation}\label{eq:K0+propto+K}
K_0 = M^{-2/3} K.
\end{equation}
Let us note that the Carter constant has the advantage of being independent of the mass, and
consequently it has meaning even in the limit when $M \to 0$;
as it appear in the 
equations of geodesic motion (\ref{eq:ellr-geod})-(\ref{eq:elltheta-geod}).

For a \emph{real vector} $k^a$ orthogonal to 
$\ell^a$ and parallel propagated along the geodesic
one has the following theorem\cite{Chandrasekhar:1985kt}:
\begin{teo}\label{teo:teo+k0}
If $\ell^a$ is a null affine geodesic vector and $k^a$ \underline{a real vector} orthogonal to 
$\ell^a$ and parallel propagated along the geodesic then the constant 
$K_0$ of theorem (\ref{teo:K0}) is proportional to the square of the modulus of the constant 
$\mathbb{K}_{(k)}$ along null geodesic.
The proportionality is given by
\begin{equation}
\mathbb{K}_{(k)}\overline{\mathbb{K}}_{(k)} = 
\left| \mathbb{K}_{(k)} \right|^2 = -\left(k^a k_a \right) K_0.
\end{equation}	
\end{teo}
It should be remarked that there is an obvious typo in the last
line of equation (34) of page 324 of reference \cite{Chandrasekhar:1985kt};
which it can be deduced from the penultimate line of the same
equation; that is the last line has an extra factor of 2 and 
using our notation for the vectors it should
just be 
$\mathbb{K}_{(k)}\overline{\mathbb{K}}_{(k)}=-2 |\Psi_2|^{-2/3} (\ell \cdot \tilde{m}) (\ell \cdot \bar{\tilde{m}}) |k|^2$.

In our applications we have considered $\mathbb{K}_{(m)}$
where $m^a$ is a complex null vector and so the above theorem is not directly
applicable to $\mathbb{K}_{(m)}$ since it assumes $k^a$ to be a real vector.
However, we can make use of it by noting that one can always write
\begin{equation}\label{eq:m+in+complex+form}
m^a = \frac{1}{\sqrt{2}}\left( x^a + i y^a \right);
\end{equation}
with $x^a$ and $y^a$ two spacelike vectors orthogonal to $\ell^a$ satisfying:
$x^a x_a = y^a y_a = -1$ and $x^a \ell_a = y^a \ell_a = 0$.
Then, since we have required $m^a$ to be parallel transported along the null 
geodesic both $x^a$ and $y^a$ are parallel transported as well.
From theorem \eqref{teo+1} this implies 
that
\begin{equation}
\mathbb{K}_{(m)} = \mathbb{K}_{(x/\sqrt{2})} + i \mathbb{K}_{(y/\sqrt{2})} \neq 0;
\end{equation}
while
\begin{equation}
\mathbb{K}_{(\bar{m})} = \mathbb{K}_{(x/\sqrt{2})} - i \mathbb{K}_{(y/\sqrt{2})} = 0.
\end{equation}

Then, from theorem \eqref{teo:teo+k0} one arrives at:
\begin{corollary}
Let $\ell^a$ is a null affine geodesic vector and $m^a$ \underline{a complex null vector},
of the form \eqref{eq:m+in+complex+form}, orthogonal to $\ell^a$ and parallel propagated along 
the geodesic. 
Then the constant $K_0$ of theorem (\ref{teo:K0}) is proportional to the square of the modulus 
of the constant $\mathbb{K}_{(m)}$ along the null geodesic.
The proportionality in such case is given by
\begin{equation}\label{eq:Kgorda}
\begin{split}
\mathbb{K}_{(m)} \overline{\mathbb{K}}_{(m)} 
= 2 K_0 
= 2 M^{-2/3} K
.
\end{split}
\end{equation}
\end{corollary}
This is a very simple expression relating $K$ with the modulus of $\mathbb{K}_{(m)}$.

In what follows we will avoid the subindex in $\mathbb{K}_{(m)}$ and we will 
just write $\mathbb{K}$.

\subsection{The exact computation of $\mathbb{K}$}

The constant $\mathbb{K}$ is given from equation (\ref{eq:theorem-K}) by
\begin{equation}\label{eq:K}
\mathbb{K} = 2\left[(\ell^a \tilde{\ell}_a )(m^a \tilde{n}_a) - 
(\ell^a \tilde{m}_a)(m^a \bar{\tilde{m}}_a) \right] \tilde{\Psi}_2^{-1/3}
;
\end{equation}
where the intervening contractions are:

\begin{equation}\label{eq:ell-ellprinc-asympt}
\begin{split}
\ell_a \tilde{\ell}^a = \frac{ E (r_o^2 + a^2) - a L_z - \sqrt{\mathcal{R}_o}}{\Delta_o} 
,
\end{split}
\end{equation}
\begin{equation}\label{eq:m-nprinc-asympt}
\begin{split}
m_a \tilde{n}^a =& 
\Big( i \cos\big(\frac{\pi}{2} - \delta_{z} \big) 
\cos\big( \frac{\pi}{2} - \alpha_{x} \big)
- \sin\big( \frac{\pi}{2} - \alpha_{x} \big) \Big) 
\\
&
\frac{\sqrt{\Delta_o} a \sin(\theta_o)}
{N_{\mathsf{y}} 2\sqrt{2} \sqrt{\Sigma_o} \sqrt{1 - \varPhi_o}}
\Big( 
\varPhi_o 
+
\frac{\sqrt{\mathcal{R}_\text{cm} }}{ \Sigma_o}
\Big)
\\
&+  
\Big( i \cos\big(\frac{\pi}{2} - \delta_{z} \big) 
\sin\big( \frac{\pi}{2} - \alpha_{x} \big) 
+ \cos\big( \frac{\pi}{2} - \alpha_{x} \big) \Big) \\
&
\frac{\sqrt{\Delta_o}}{N_{\mathsf{y}} 2 \sqrt{2} \Sigma_o}
\Big(
\sqrt{\mathcal{R}_\text{cm}}
-
\frac{\varPhi_o\, a^2 \sin(\theta_o)^2}{1 - \varPhi_o} 
\Big)
,
\end{split}
\end{equation}
\begin{equation}\label{eq:ell-mprinc-asympt}
\begin{split}
\ell_a \tilde{m}^a =& \frac{1}{\sqrt{2} \mathfrak{r}_o}
\Big(
(-1)
\left( \pm \sqrt{\varTheta_o} \right) + i \big( aE \sin(\theta_o) - \frac{L_z}{\sin(\theta_o)}\big)
\Big) 
,
\end{split}
\end{equation}
\begin{equation}\label{eq:m-mbarprinc-asympt}
\begin{split}
m_a \bar{\tilde{m}}^a =& 
\Big( \cos\big(\frac{\pi}{2} - \delta_{z} \big) 
\cos\big( \frac{\pi}{2} - \alpha_{x} \big)
+ i \sin\big( \frac{\pi}{2} - \alpha_{x} \big) 
\Big) 
\\
&
\frac{\sqrt{\Delta_o}}{ 2 N_{\mathsf{y}}}
\Big( 
\frac{1}{ \bar{\mathfrak{r}}_o }
\sqrt{\frac{\mathcal{R}_\text{cm} }{\Sigma_o (1 - \varPhi_o)} }
\Big)
\\
&+  
\Big(
- \cos\big(\frac{\pi}{2} - \delta_{z} \big) 
\sin\big( \frac{\pi}{2} - \alpha_{x} \big) 
+ i \cos\big( \frac{\pi}{2} - \alpha_{x} \big) 
\Big) \\
&
\frac{\sqrt{\Delta_o}}{2 N_{\mathsf{y}}}
\Big(
\frac{\varPhi_o a \sin(\theta_o)}{\bar{\mathfrak{r}}_o (1 - \varPhi_o) } 
\Big)
\\
&+  i \sin\big(\frac{\pi}{2} - \delta_{z} \big) 
\frac{\sqrt{\Sigma_o}}{2 \bar{\mathfrak{r}}_o } 
.
\end{split}
\end{equation}

These allow for the exact computation of the constant  $\mathbb{K}$
in terms of the position of the observer and the angles chosen, in the sphere of directions,
for the observed null geodesic.

\subsection{Computation of $\mathbb{K}$ and the curvature scalar $\Psi_0$ for a distant observer}\label{subsec:Psi-0}
Let us study the explicit expression of $\mathbb{K}$ when the observer lies in a region
where $\frac{M}{r_o} <<1$ and $\frac{a}{r_o} <<1$: so that it is natural to
study each expression in terms of orders
$\mathscr{O}\left( \frac{M}{r_o} \right)$ and 
$\mathscr{O}\left( \frac{a}{r_o} \right)$ while the constant of motion
$K$ and $L_z$ are kept constant.
The expansion in terms of these orders gives
\begin{align}
\ell^a \tilde{\ell}_a =& 
\frac{K }{2 E r_o^2 } + \mathscr{O}(\frac{K M}{r_o^3})
;
\\
m_a \tilde{n}^a =&
\frac{\alpha_x - a\sin(\theta_o) + i\delta_z}{2\sqrt{2}}\bigg( 1+ \mathscr{O}\Big( \frac{M}{r_o}\Big) \bigg),
\\
\begin{split}
\ell^a \tilde{m}_a =& -\frac{1}{\sqrt{2}r_o}\Bigg[
\pm \sqrt{K - \left( \frac{L_z}{\sin(\theta_o)} - \sqrt{K_{o_\text{cm}}} \right)^2}  
\\
&  - i\left(\sqrt{K_{o_\text{cm}}} - \frac{L_z}{\sin(\theta_o)}\right)\Bigg]
\bigg(
1	+ \mathscr{O}\Big( \frac{a}{r_o} \Big)
\bigg)
,
\end{split}
\\
m^a \bar{\tilde{m}}_a =& i 
\bigg(
1	+	\mathscr{O}\Big( \frac{M}{r_o} \Big) 
\bigg)
,
\end{align}
and then
\begin{equation}
\begin{split}
\mathbb{K} =& 
- 2 (\ell_a \tilde{m}^a )(m^a \bar{\tilde{m}}_a) \tilde{\Psi}_2^{-1/3} 
+ \mathscr{O} \Big( \frac{aK}{r_o^2} M^{-1/3} \Big)
\\
=&
-\frac{i \sqrt{2}}{ M^{1/3}}\left[
\delta_z r_o
- 
i\left(\sqrt{K_{o_\text{cm}}} + \alpha_x r_o \right)\right]
\bigg(
1	+	\mathscr{O}\Big( \frac{M}{r_o} \Big) 
\bigg)
\\
& + \mathscr{O} \Big( \frac{aK}{r_o^2} M^{-1/3} \Big)
.
\end{split}
\end{equation}

It is interesting to see that this expression satisfies equation \eqref{eq:Kgorda}.

In this we arrive at the main result of this section,
since we can now express the curvature scalar $\Psi_0$; 
which, taking into account \eqref{eq:Psi0+theorem}, is: 
\begin{equation}
\begin{split}\label{eq:psi0exacto}
\Psi_0 (r, \theta )
=& \frac{3M}{\left( r - ia \cos(\theta)\right)^5}
\left[ \delta_z r_o - i \left( \sqrt{K_{o_\text{cm}}} + \alpha_x r_o \right) \right]^2 
\\
&  \bigg(
1	+	\mathscr{O}\Big( \frac{M}{r_o} \Big) 
\bigg)
+ \mathscr{O} \Big( \frac{aK^{3/2} }{r_o^2} M^{-2/3} \Big)
.
\end{split}
\end{equation}

Let us note that all the coordinate dependence of $\Psi_{0}$ appears
in the denominator of the first factor.

\section{Numerical calculations of the optical scalars}\label{sec:numerical}

In this section, just to give an application of the efficient expressions,
we present the calculation of the differences from the exact numerical calculation
of the optical scalars, as explained in section \ref{sec:GeodDev+WeakLens},
with the thin lens first order weak lens expressions given by
\eqref{eq:g1masig2} and \eqref{eq:integrated-weyl}; for
a particular case of a not moving Kerr lens.

\subsection{Thin and weak lensing of Kerr spacetime}

In this subsection we describe the trajectory of the photons in terms
of the observed angles, as used in the numerical calculation of the
weak lens effects at first order.

We recall that in such approximation, the convergence $\kappa$ vanishes since Kerr is Ricci 
flat and therefore the only-non trivial optical scalar is the shear which is given by:
\begin{equation}\label{eq:shear+Kerr+WL}
\begin{split}
\gamma_{1} + i\gamma_2 =& 3M 
\frac{d_l d_{ls}}{d_s }  
\left[ \delta_z r_o - i \left( \sqrt{K_{o_\text{cm}}} + \alpha_x r_o \right) \right]^2
\\
& \int_{0}^{\lambda_s} \frac{d\lambda'}{\left(r - ia\cos(\theta) \right)^5}.
\end{split}
\end{equation}

The above integral is prescribed along null geodesics of Kerr, however, usually the integration is done 
along the straight path that photons would follow on the flat background.

Let us denote with $(x,y,z)$ the Cartesian coordinates used by the observer
to describe the situation, and with $(x',y',z')$ the Cartesian coordinates
obtained from the Boyer-Lindquist coordinate system in the limit for
the mass going to zero.
In the calculation of weak lens effects at first order of the curvature,
these two coordinate systems are related by a rotation and a translation.
We would like to express the primed systems in terms of the un-primed one.
First, let us consider the rotation around the $x$ axis so that
\begin{align}
x'' &= x , \\
y'' &= y \cos(\iota) + z \sin(\iota) , \\
z'' &= -y \sin(\iota) + z \cos(\iota)
.
\end{align} 
Then, the primed system is obtained by a translation
\begin{align}
x' &= x'' , \\
y' &= y'' - d_l \cos(\iota) , \\
z' &= z'' + d_l \sin(\iota)
; 
\end{align}
where one can note that at the origin of the $(x,y,z)$ system,
that is at the position of the observer, one has
$(x''~=~0,y''~=~0,z''=0)$
and
$(x'=0,y'=-d_l \cos(\iota),z'=d_l \sin(\iota))$.
Then, we have
\begin{align}
x' &= x , \\
y' &= y \cos(\iota) + z \sin(\iota)  - d_l \cos(\iota) , \\
z' &= -y \sin(\iota) + z \cos(\iota) + d_l \sin(\iota)
;
\end{align}
where
\begin{align}
x &= r_o \alpha_x , \\
y &= \lambda , \\
z &= r_o \delta_z .
\end{align}
These relations are used to calculate $r$ from
\begin{equation}
\begin{split}
r^2 =& \frac{1}{2}
\left(
x'^2+y'^2+z'^2 - a^2 \right. \\
&+  \left. \sqrt{(x'^2+y'^2+z'^2 - a^2)^2 + 4 z'^2 a^2}
\right)
,
\end{split}
\end{equation}
and $\cos(\theta)$ from
\begin{equation}
\cos(\theta) = \frac{z'}{r}
.
\end{equation}

\subsection{Comparison of the first order weak lens approximation with the exact numerical
	calculation}

The next graphs show the result of the computation of the shear (gamma)
using the expressions for weak lens effects in first order, along with its
comparison with the exact calculation on the true geodesic trajectory
and geodesic deviation equation.
The points of observation, in the plane of the image, 
have been chosen to be on a straight line at 60 degrees from the
line aligned with $z$ axis of the observer frame.
The minimal $\alpha$ angle ($x$ observational axis) we have considered,
with respect to the center of the black hole, is of about 0.24arcsec;
which corresponds to a $\delta$ angle ($z$ observational axis)
of about 0.13arcsec.
The geometric values of the black hole are taken from the 
M87, also known as Virgo A or NGC 4486.
It is one of the most massive galaxies close to us, located at a distance of 
16.7 million parsecs (Mpc) with a jet that extends at least 1500 parsecs.
The total mass is about $2.4 \times 10^{12}$ solar masses.
The black hole we are interested in is the 
supermassive black hole at the center of M87,
with mass of $M=6.7 \times 10^{9}$ solar masses.
The angular momentum of this black holes,
according to \cite{Feng:2017vba},
can be characterized by a Kerr parameter of 
$a = 0.98 M$; which we have used in our calculations.
For the sources we will assume them to be located at 40Mpc.
For our calculations we have also assumed the angular momentum to be tilted
an arbitrary  angle $\iota = \frac{\pi}{6}$ towards the observer;
which does not coincide with the estimated inclination of the supermassive
black hole in M87.

The exact numerical calculation is carried out using a Runge-Kutta routine\cite{rksuite-90}
using the high order (7,8) pair.
The first order weak lens shear was calculated using a flat background geodesic,
dividing the region of quadrature in three: before the lens, across the lens,
and after the lens. The regions where adjusted to optimize the results
of each quadrature, and the sum of them. The quadratures were calculated
using a 7th order routine, although a Clenshaw-Curtis type method\cite{Clenshaw60} with
Chebyshev approximation was also tested.
\begin{figure}
	\centering
	\includegraphics[clip,width=0.49\textwidth]{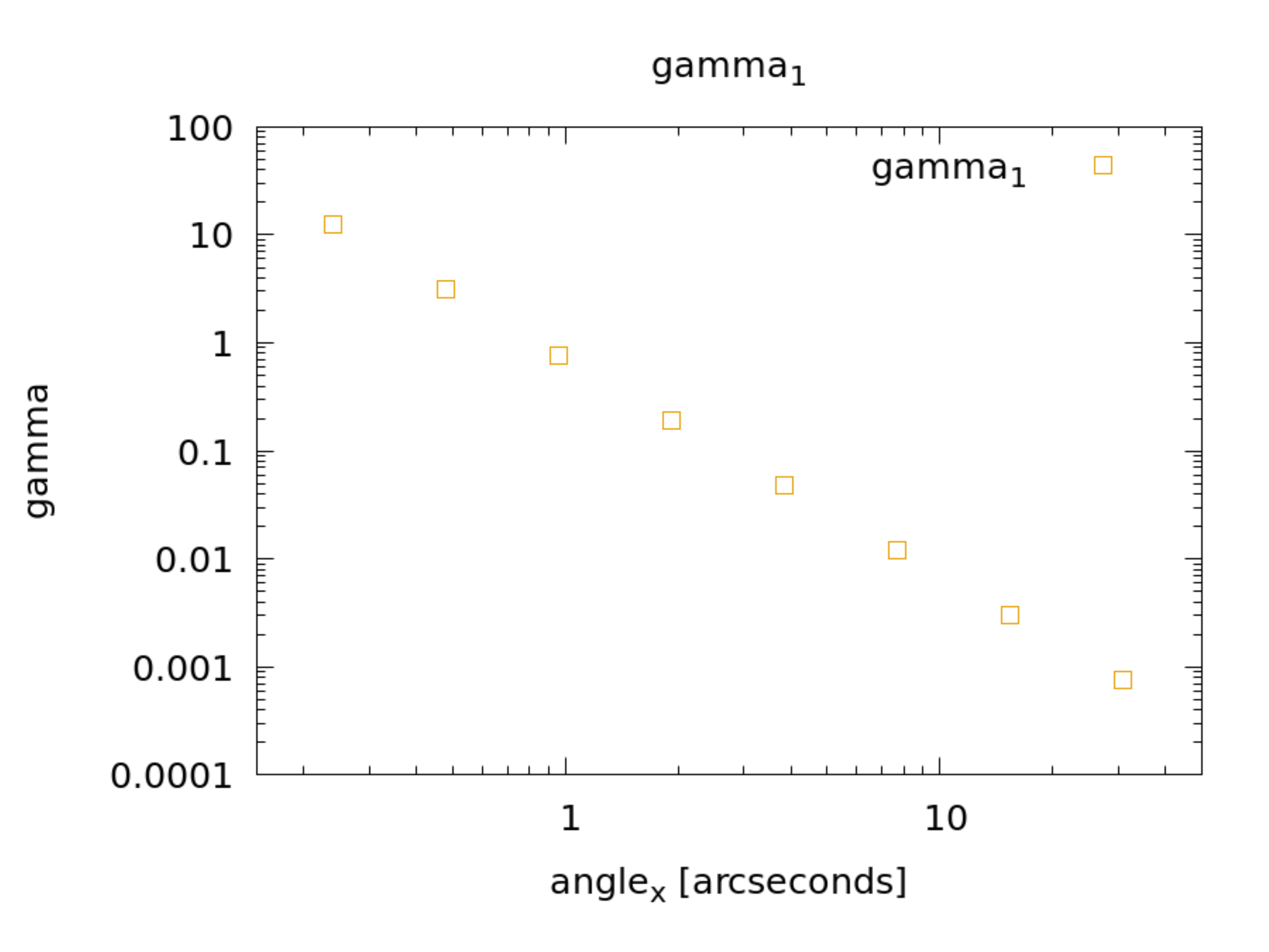}
	\includegraphics[clip,width=0.49\textwidth]{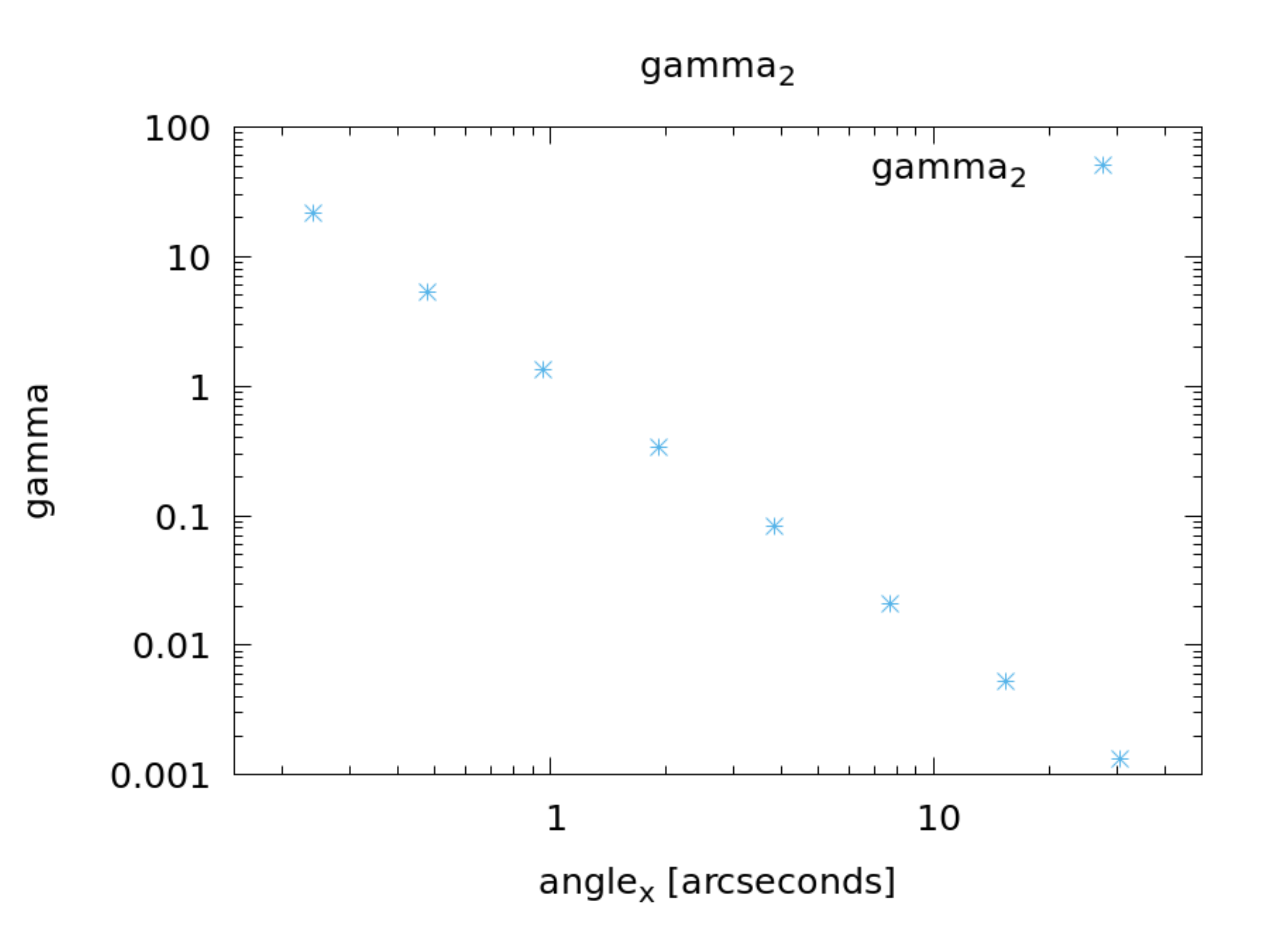}
	\caption{Values of $\gamma_1$ and $\gamma_2$ for a big range
		of angles.
		The horizontal axis is the $x$ observational axis 
		measured as angles in units of arcseconds.	
	}\label{fig:gammas}
\end{figure}

\begin{figure}
	\centering
	\includegraphics[clip,width=0.49\textwidth]{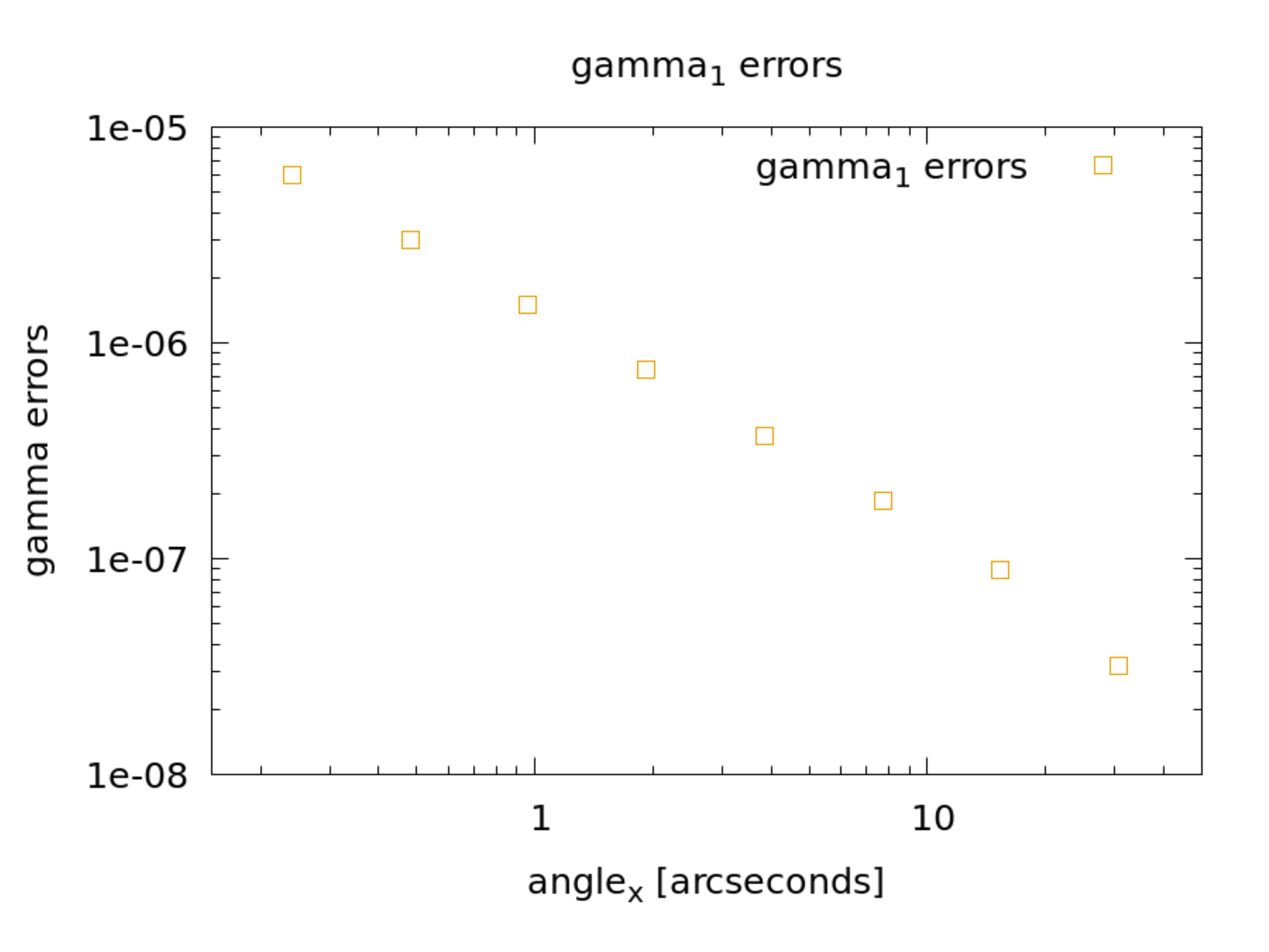}
	\includegraphics[clip,width=0.49\textwidth]{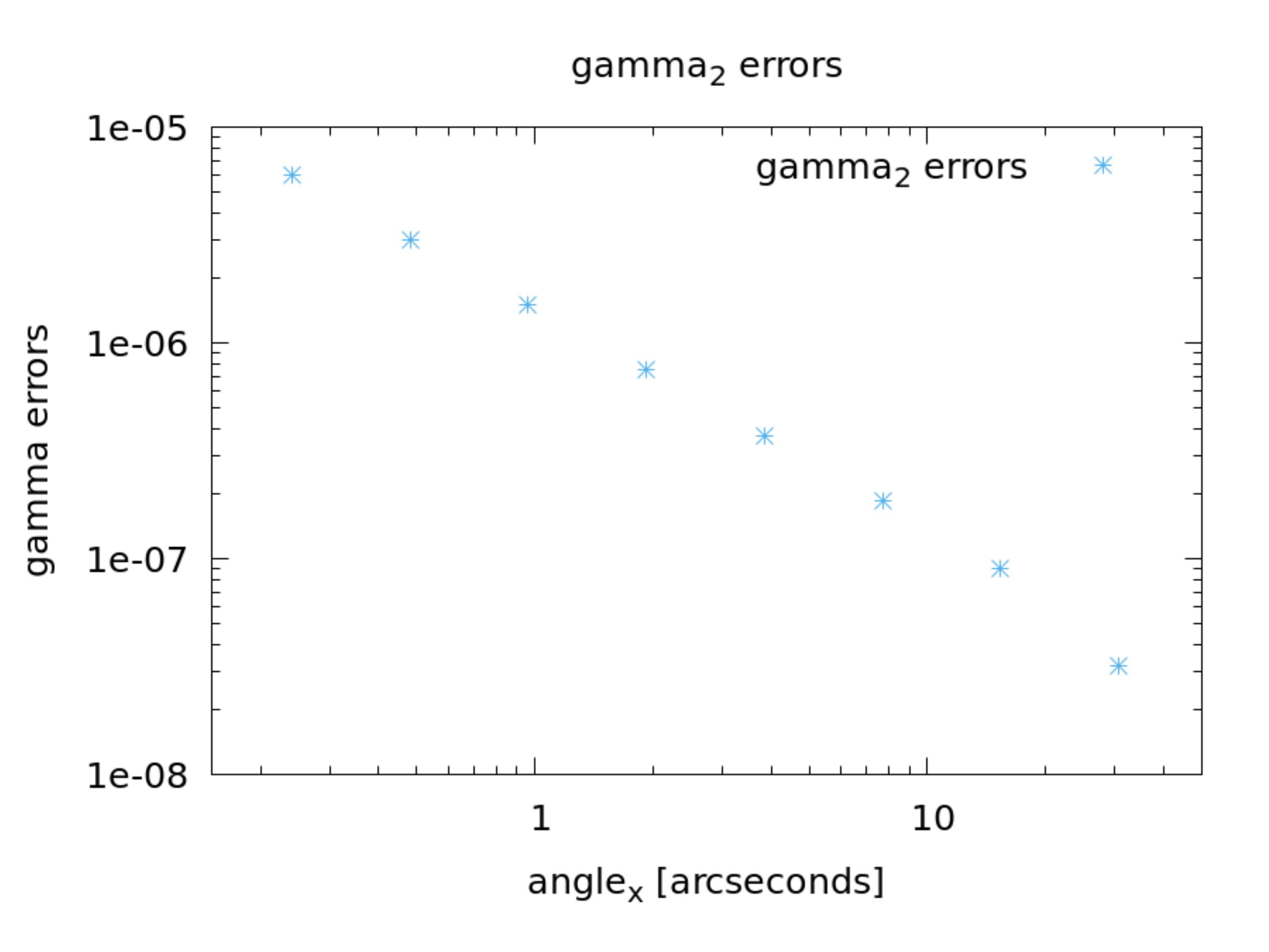}
	\caption{Relative errors from the exact numerically
		calculated values of $\gamma_1$ and $\gamma_2$.
		The horizontal axis is the $x$ observational axis
		measured as angles in units of arcseconds.	
	}\label{fig:errors-gammas}
\end{figure}
In figures \ref{fig:gammas} and \ref{fig:errors-gammas} we show the values 
of $\gamma_1$ and $\gamma_2$, calculated in the thin lens approximation,
along with the corresponding relative errors with
respect to the numerical calculation of the exact equations with
high precision.

As expected in this range of observational angles, it can be seen that 
there is an impressive good agreement between
the first order weak lens calculations and the exact calculations.

In a separate article we plan to make a detailed study of
all maps of the optical scalars, explaining extensively
how to improve at each instance for efficient calculations
and control errors.

\subsection{Optical scalars maps for weak lensing in Kerr spacetime}\label{subsec:Shear+Maps}

\begin{figure}
	\centering
	\includegraphics[clip,width=0.5\textwidth]{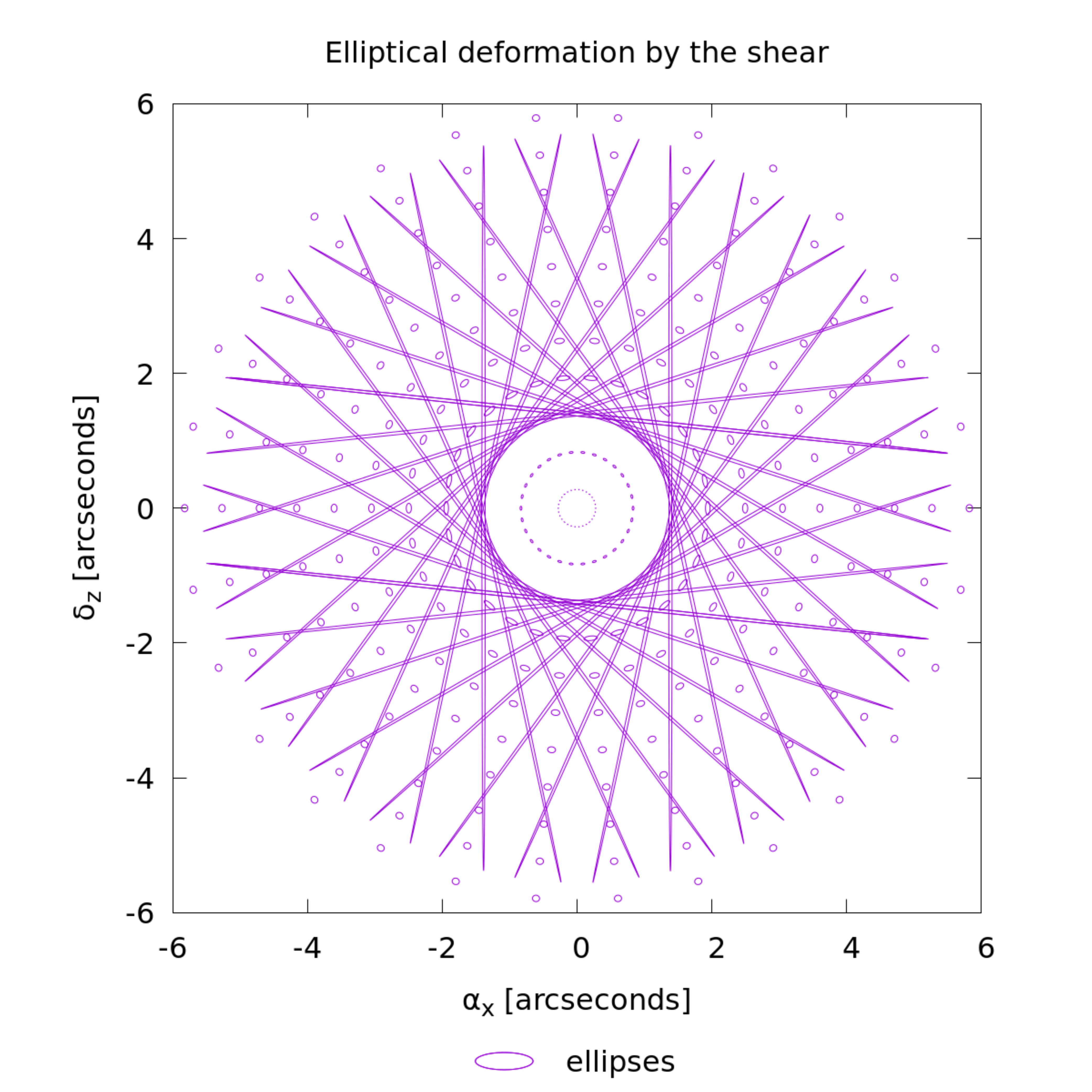}
	\caption{Elliptical deformation.
		This graph shows distorted ellipses corresponding to circular sources behind 
		the lens distributed in a polar grid. 
		The elliptical deformations increase from the exterior towards the center;
		and it is dominated by the huge deformation occurring near the Einstein ring.
		Within the inner region it is notorious the demagnification.
	}
	\label{fig:elliptical-deformation}
\end{figure}

\begin{figure}
	\centering
	\includegraphics[clip,width=0.5\textwidth]{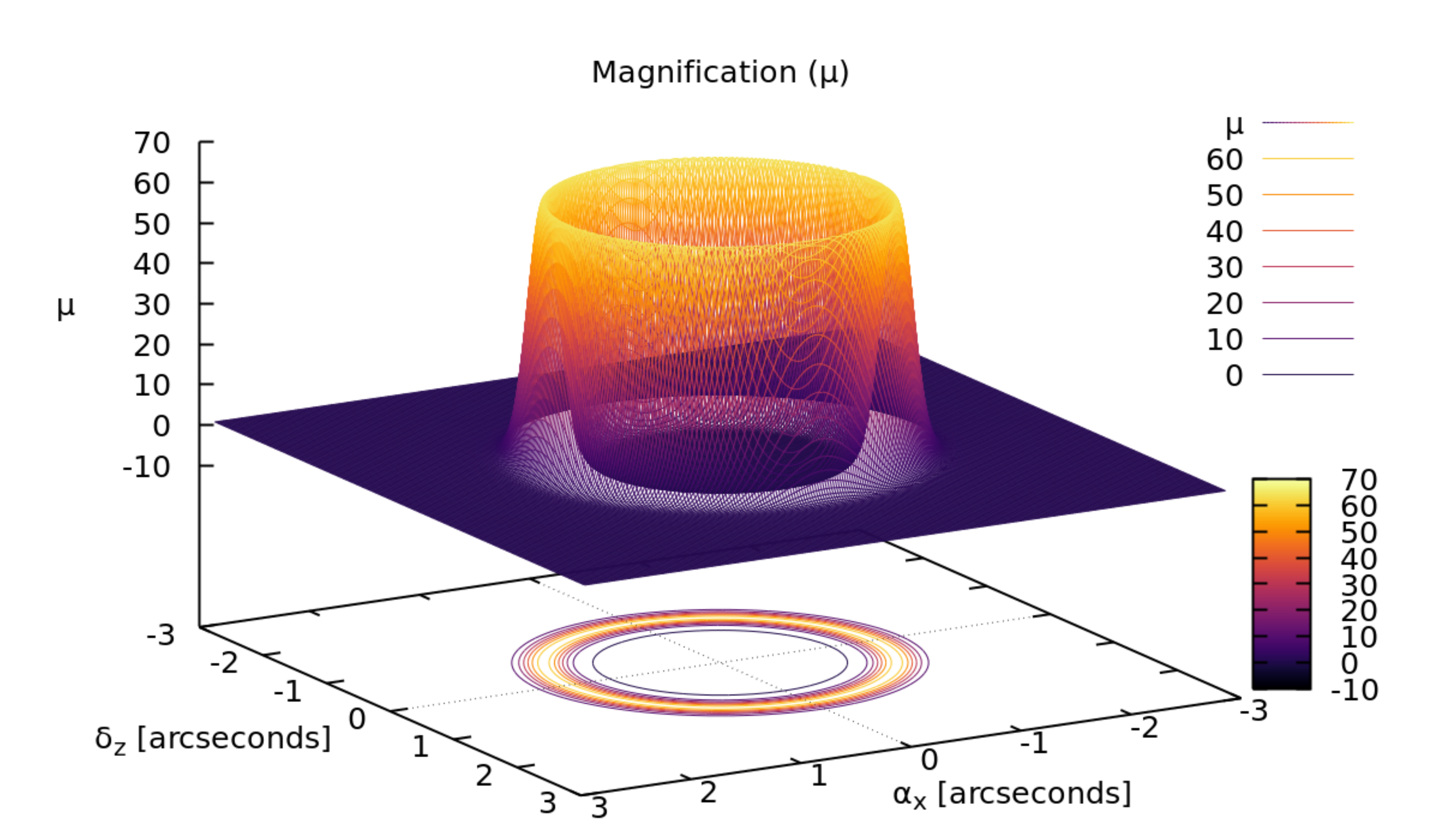}
	\caption{Magnification. 
		This graph shows the shape of Einstein ring at the scales explored. 	
	}
	\label{fig:magnification}
\end{figure}

In figure \ref{fig:elliptical-deformation} we show the ellipticity map for the same M87 system
in a range of angles in which the Einstein ring appears.
Figure \ref{fig:magnification} shows a 3d graph of the numerical calculation of the magnification
map for the M87 system.
For these parameters and this range of angles, one can not notice the effects 
due to the spinning of the black hole;
due to the fact that for this combination of lens and source distances, the Einstein
ring is located in a region where the rotational effects are small.

In all the graphs we have presented we have chosen the range of angles that include
the divergence of the magnification; which is important for the weak lens effects.
Since the trajectories of the photons close to this characteristic angle,
have minimum distances to the black hole which are large compared with its mass
scale, when performing the same calculations with different choices of the
angular parameter, give almost the same result.
Some calculations are shown in Appendix \ref{sec:app-adiferentes}.

\section{Final comments}

This article presents a detailed discussion of gravitational 
lensing in Kerr spacetime. 
The main focus is the calculation of the optical scalars appearing
in lens effects, and therefore the treatment 
is centered on the geodesic deviation equations.

The main result that we present is a surprisingly simple and compact
expression for the curvature component $\Psi_{0}$, which appears in the
optical scalar computation,
as shown in equation \eqref{eq:Psi0+theorem} of section \ref{sec:lensing-Kerr}.
This should be compared with the lengthy expressions appearing in other works,
as for example in equations 
(A.19) and (A.20), with (A.21),
used in reference \cite{James:2015yla}
in the numerical code Double Negative Gravitational Renderer (DNGR),
which were used in the calculations of images of a Hollywood movie;
see also reference \cite{Pineault1977ApJ}.
Our expressions show a the notorious simplification providing a great
improvement in the efficiency of computation of the lens effects.

We arrived at these new results by using all the power of the
hidden symmetries of vacuum type-D spacetimes\cite{Chandrasekhar:1985kt,Walker:1970un}.
One important aspect is the introduction of a natural frame for the photon
trajectory and its relation with a principal null tetrad frame.
Another main ingredient is the explicit calculation of the constants
of motion in terms of the angles, as seen by an observer;
which allows for an immediate application of our expressions
to observational works.

We have also presented as an application a numerical study of the systems of equations
in the case of weak lensing for a lens with mass, spin and distances corresponding to 
those of M87, but with an arbitrary inclination angle of $\iota = \frac{\pi}{6}$,
as an illustrative example for our study.
The graphs above, (figures \ref{fig:elliptical-deformation} and 
\ref{fig:magnification}), show the ellipticity and magnifications maps in a range of
angles in which the outer caustic is noticeable; although
the angular momentum contribution from the black hole is not detectable;
since the distance chosen for the sources sets the caustics rather far from
the central black hole.

Our result also is a contribution for the theoretical understanding
of the gravitational lens problem for black holes with angular momentum;
since we have proved, when comparing the first order weak lens model with
the exact calculation, that the simple model based on calculation on
a straight geodesic of the natural flat background, gives excellent
results; and therefore one can avoid the use of complicated 
bending angle models based on 3-dimensional set of lines\cite{Renzini:2017bqg}.
Let us emphasize that our discussion does not use the notion of bending angle
at all; since instead we use the exact geodesic deviation equations.

We have also introduced a new way to refer to what one considers
the direction, in the sky, that points to the center of a black hole
with angular momentum.
Of course, at first sight the notion of center of the black-hole is puzzling,
among other things this is due to the fact that the singularity can be interpreted as a 
ring which is hidden to the sight of the observer 
and since the shadow has no center of symmetry.
The standard reference direction has been to use the one
associated to an orthogonal frame constructed from the 
coordinate directions\cite{Chandrasekhar:1985kt}
at the position of the observer. 
Instead, our criteria for making a choice 
has been to take the spatial directions with respect to the 
direction of the null geodesic passing by the observer and pertaining to the 
center of mass null geodesic congruence as developed in \cite{Arganiaraz19a},
for any assumed position of the observer.

The equations presented here will be useful for
detailed calculations of gravitational lens effects 
in the strong regime of black holes with angular
momentum, as those studied by the 
Event Horizon Telescope Collaboration\cite{Akiyama:2019cqa,Akiyama:2019brx,Akiyama:2019sww,
	Akiyama:2019bqs,Akiyama:2019fyp,Akiyama:2019eap};
which we plan to carryout in the near future.

\subsection*{Acknowledgments}

We acknowledge support from CONICET, SeCyT-UNC and Foncyt.

\appendix

\section*{Appendices}

\section{Further comments on the center of mass function 
$K$}\label{ap:Center+mass+Kfunction}

The problem discussed in \cite{Arganiaraz19a} is related to the idea
that one would like to request a congruence of null geodesics, as described by
\begin{equation}\tag{\ref{eq:dele-general}}
\begin{split}
\ell_a =& 
\,E dt_a 
- 
\frac{\pm 	\sqrt{\mathcal{R} }
}{\Delta} dr_a 
- \big( \pm \sqrt{ \Theta }   \big)  d\theta_a 
- L_z d\phi_a
;
\end{split}
\end{equation}
to generate a null hypersurface with selected properties.
There, we chose the condition that the null hypersurfaces should coincide at
future null infinity with the center of mass\cite{Moreschi04,Gallo:2014jda}
sections; which implies $L_z=0$ and the asymptotic condition
$K= a^2 \sin^2(\theta^*)$, where $\theta^*$ is the value of the angular
coordinate $\theta$ at the point in the asymptotic sphere
reached by the future null geodesic.
Then, without loss of generality we take $E=1$ for the congruence, 
and one has the condition that the exterior derivative
of $\ell_a$ should vanish; which provides a partial differential equation
for $K$, which is now understood as a function in the interior
of the spacetime. That is, for any fix point, with coordinates $(r,\theta)$,
 in the interior of the spacetime,
there passes just one null geodesic, belonging the the center of mass family,
which has a particular value of $K_\text{cm}$. The condition $d\ell = 0$,
provides with an equation that determines $K_\text{cm}(r,\theta)$
with the boundary condition $K_\text{cm}= a^2 \sin^2(\theta^*)$.

It is possible to calculate $K_\text{cm}$ with any desired precision\cite{Arganiaraz19a}.
In these discussions it appears naturally the comparison with
the limiting calculation of these null congruences as one approaches
Minkowski spacetime, where the limiting value of $K_\text{cm}$ is given by:
\begin{equation}\tag{\ref{eq:ktilde}}
\tilde K(r,\theta)
=
\frac{(r^2 + a^2)\, a^2 \sin(\theta)^2 }{r^2 + a^2 \sin(\theta)^2}
.
\end{equation}
From this we define $\tilde{k}$ from the relation
\begin{equation}\tag{\ref{eq:kchica2}}
K_\text{cm}(r,\theta) = \tilde K - \tilde k(r,\theta)^2 ;
\end{equation}
where $\tilde K$ is the function $K_\text{cm}$ in the limit for the mass
$M\rightarrow 0$ shown in \eqref{eq:ktilde}.


\section{Restricted Lorentz transformations}\label{ap:hom-Lorentz-trans}

We mention in this appendix several relations that appear in the transformations,
however not all of them are used in our calculations.

\subsection{General transformations}
The restricted Lorentz group usually denoted by $SO^+(3,1)$ has the property that 
it can be expressed as the product of three subgroups.
These subgroups acting on the null tetrad can be parametrized by means of 
two complex functions, namely $\varLambda$ and $\varGamma$ and by two real
ones, namely $Z$ and $s$ in the following way\cite{Prior77}:
\begin{align}
\ell^a &\to \ell^a , \label{eq:subgroupI+ell}\\
m^a &\to m^a + \varGamma \ell^a, \\
n^a &\to n^a + \bar{\varGamma} m^a + \varGamma \bar{m}^a + \varGamma \bar{\varGamma} \ell^a;
\label{eq:subgroupI+n}
\end{align}
\begin{align}
n^a &\to n^a \label{eq:subgroupII+n},\\
m^a &\to m^a + \varLambda n^a, \\
\ell^a &\to \ell^a + \bar{\varLambda} m^a + \varLambda \bar{m}^a + \varLambda \bar{\varLambda} n^a;
\label{eq:subgroupII+ell}
\end{align}
\begin{align}
\ell^a &\to Z\ell^a , \label{eq:subgroupIII+ell}\\
m^a &\to m^a e^{is}, \\
n^a &\to \frac{1}{Z} n^a \label{eq:subgroupIII+n}.
\end{align}

The transformation (\ref{eq:ell-Lorentz})-(\ref{eq:n-Lorentz}) is obtained from:
\begin{align}
n^a =& \tilde{n}^a ,\\
m^a =&\tilde{m}^a + \varLambda \tilde{n}^a, \\
\ell^a =& \tilde{\ell}^a + \bar{\varLambda} \tilde{m}^a + \varLambda \bar{\tilde{m}}^a 
+ \varLambda \bar{\varLambda} \tilde{n}^a;
\end{align}
in which we have used the second subgroup listed above (\ref{eq:subgroupII+n})-(\ref{eq:subgroupII+ell}).
Then let us apply the third subgroup, namely equations (\ref{eq:subgroupIII+ell})-(\ref{eq:subgroupIII+n}),
this yields
\begin{align}
n^a =& \frac{1}{Z} \tilde{n}^a ,\\
m^a =& \Big( \tilde{m}^a + \varLambda \tilde{n}^a \Big) e^{is}, \\
\ell^a =& Z \Big( \tilde{\ell}^a + \bar{\varLambda} \tilde{m}^a + \varLambda \bar{\tilde{m}}^a 
+ \varLambda \bar{\varLambda} \tilde{n}^a \Big);
\end{align}
and finally let us consider the first group (\ref{eq:subgroupI+ell})-(\ref{eq:subgroupI+n})
which gives:
\begin{align}
\ell^a =& Z \Big( \tilde{\ell}^a + \bar{\varLambda} \tilde{m}^a + \varLambda \bar{\tilde{m}}^a 
+ \varLambda \bar{\varLambda} \tilde{n}^a \Big)
; 
\\
\begin{split}
m^a =& \Big( \tilde{m}^a + \varLambda \tilde{n}^a \Big) e^{is}
\\
&+ \varGamma
Z \Big( \tilde{\ell}^a + \bar{\varLambda} \tilde{m}^a + \varLambda \bar{\tilde{m}}^a 
+ \varLambda \bar{\varLambda} \tilde{n}^a \Big), 
\end{split}
\\
\begin{split}
n^a =& \frac{1}{Z} \tilde{n}^a + \bar{\varGamma}\Big( \tilde{m}^a + \varLambda \tilde{n}^a \Big) e^{is}
+ \varGamma\Big( \bar{\tilde{m}}^a + \bar{\varLambda} \tilde{n}^a \Big) e^{-is}
\\
& + \varGamma \bar{\varGamma} Z
\Big( \tilde{\ell}^a + \bar{\varLambda} \tilde{m}^a + \varLambda \bar{\tilde{m}}^a 
+ \varLambda \bar{\varLambda} \tilde{n}^a \Big)
.
\end{split}
\end{align}
Is easy to see that the above equation simplify to the set 
\eqref{eq:ell-Lorentz} - \eqref{eq:n-Lorentz}.

\subsection{Expressions for the functions $Z$ and $\varLambda$}

From equation \eqref{eq:ell-Lorentz}, we observe that we can compute the 
$Z$ along a null geodesic in the following way:
\begin{equation}
\begin{split}
Z = \ell^a \tilde{n}_a =&
\frac{\Delta}{2 \Sigma} \dot{t} + \frac{1}{2}\dot{r} - \frac{a \Delta \sin(\theta)^2}{2 \Sigma} \dot{\phi} 
\\
=& 
\frac{1}{2\Sigma}\Bigg( E(r^2 + a^2) - a L_z \pm \sqrt{\mathcal{R}} \Bigg)
\\
=&
\frac{1}{2\Sigma}\Bigg( \sqrt{\mathcal{R} + K \Delta} \pm \sqrt{\mathcal{R}} \Bigg).
\end{split}
\end{equation}

Similarly, from equation \eqref{eq:ell-Lorentz} an expression for $\varLambda$ along any null 
geodesic is given by:
\begin{equation}
\begin{split}
\varLambda =& - \frac{\ell^a \tilde{m}_a}{Z} 
\\
=&
-\frac{1}{\sqrt{2}\mathfrak{r} Z}\Bigg( ia\sin(\theta) \dot{t} 
- \bigg(
\Sigma \dot{\theta} 
+ i(r^2 + a^2)\sin(\theta) \dot{\phi}
\bigg) \Bigg)
\\
=&\frac{1}{\sqrt{2}\mathfrak{r} Z}\bigg( \pm \sqrt{\varTheta}
- i \sqrt{K - \varTheta}
\bigg).
\end{split}
\end{equation}

\subsection{Expressions for the functions $\varGamma$ and $e^{is}$ at the position of the observer}
Looking at equation \eqref{eq:m-Lorentz} we realize that $\varGamma$ and $e^{is}$ are given, 
respectively by
\begin{equation}
\begin{split}
\varGamma = \frac{m^a \tilde{n}_a}{Z};
\end{split}
\end{equation}
and
\begin{equation}
e^{is} = - m^a \bar{\tilde{m}}_a - \varGamma Z \bar{\varLambda} = 
- m^a \bar{\tilde{m}}_a -  m^a \tilde{n}_a \bar{\varLambda}.
\end{equation}

\section{Distortion effects produced by the optical scalars}

\subsection{Shear maps}\label{subsec:shear+maps}
Weak lensing observations only reveal information about the shear 
$\gamma_1 + i \gamma_2 = - \gamma e^{2i\vartheta}$ produced by the lens.

In absence of twist, $\omega = 0$, the effect of shear manifests very clearly
through the linear distortions of small extended sources;
for example in the case of
small circular sources, 
the resulting images are small ellipses with is major axis rotated an 
angle $\varphi = \vartheta$ respect to the $x$-axis associated to the coordinates 
$(\alpha_{x}, \delta_{z})$.
Instead, when $\omega \neq 0$ the major axis of the small ellipse is rotated an angle $\varphi \approx
\vartheta + \frac{1}{2}\frac{\omega}{\gamma}$.

If we denote by $\left(\delta \beta_x , \delta \beta_z\right)$ the components 
of $\delta \beta^a$ and by $(\delta {\alpha}_{x}, \delta {\delta}_{z})$
the components of $\delta \theta^a$ in the plane expanded by
$({\alpha}_{x}, {\delta}_{z})$:
\begin{equation}
\delta \beta^a = \begin{pmatrix}
\delta \beta_{x} \\
\delta \beta_{z}
\end{pmatrix},
\end{equation}
\begin{equation}
\delta \theta^a = \begin{pmatrix}
\delta {\alpha}_{x} \\
\delta {\delta}_{z}
\end{pmatrix};
\end{equation}
then we have 
\begin{equation}
\begin{pmatrix}
\delta \beta_{x} \\
\delta \beta_{z}
\end{pmatrix}
=
\begin{pmatrix}
1 - \kappa - \gamma_1 & -\gamma_2 -\omega \\
- \gamma_2 + \omega & 1 - \kappa + \gamma_1
\end{pmatrix}
\begin{pmatrix}
\delta {\alpha}_{x} \\
\delta {\delta}_{z}
\end{pmatrix}.
\end{equation}

Let $T= 2 (1 - \kappa)$
be the trace of the matrix and
$D~=~(1 - \kappa)^2 - \gamma^2 + \omega^2$
be the determinant of the matrix, where $\gamma^2 =\gamma_1^2 + \gamma_2^2$;
then the eigenvalues are
\begin{equation}
\lambda_+ = \frac{T}{2} + \sqrt{ \frac{T^2}{4} - D }
= 1 - \kappa + \sqrt{ \gamma^2 - \omega^2}
,
\end{equation}
and
\begin{equation}
\lambda_- = \frac{T}{2} - \sqrt{ \frac{T^2}{4} - D }
=  1 - \kappa - \sqrt{ \gamma^2 - \omega^2}
.
\end{equation}
The eigenvectors are:
\begin{align}
w_{+}^a =& 
\begin{pmatrix}
\lambda_+ - (1 - \kappa + \gamma_1) \\
- \gamma_2 + \omega
\end{pmatrix}
=
\begin{pmatrix}
\sqrt{ \gamma^2 - \omega^2} - \gamma_1 \\
- \gamma_2 + \omega
\end{pmatrix}
,
\\
w_{-}^a =& 
\begin{pmatrix}
\lambda_- - (1 - \kappa + \gamma_1) \\
- \gamma_2 + \omega
\end{pmatrix}
=
\begin{pmatrix}
-\sqrt{ \gamma^2 - \omega^2} - \gamma_1 \\
- \gamma_2 + \omega
\end{pmatrix}
,
\label{eq:eigenv+menos+I}
\end{align}
for $(- \gamma_2 + \omega) \neq 0$;
\begin{align}
w_{+}^a =& 
\begin{pmatrix}
- \gamma_2 - \omega \\
\lambda_+ - (1 - \kappa - \gamma_1) 
\end{pmatrix}
=
\begin{pmatrix}
- \gamma_2 - \omega \\
\sqrt{ \gamma^2 - \omega^2} + \gamma_1 
\end{pmatrix},
\\
w_{-}^a =& 
\begin{pmatrix}
- \gamma_2 - \omega \\
\lambda_- - (1 - \kappa - \gamma_1) 
\end{pmatrix}
=
\begin{pmatrix}
- \gamma_2 - \omega \\
-\sqrt{ \gamma^2 - \omega^2} + \gamma_1 
\end{pmatrix}
,
\label{eq:eigenv+menos+II}
\end{align}
for $(- \gamma_2 - \omega) \neq 0$;
\begin{align}
w_{+}^a =& 
\begin{pmatrix}
1 \\
0 
\end{pmatrix},
\\
w_{-}^a =& 
\begin{pmatrix}
0 \\
1  
\end{pmatrix},
\end{align}
if $\gamma_2 = \omega=0$ and $\gamma_1 < 0$;
and
\begin{align}
w_{+}^a =& 
\begin{pmatrix}
0 \\
1 
\end{pmatrix},
\\
w_{-}^a =& 
\begin{pmatrix}
1 \\
0  
\end{pmatrix}
,
\end{align}
if $\gamma_2 = \omega=0$ and $\gamma_1 > 0$.

The normalized eigenvectors for the first case $(- \gamma_2 + \omega) \neq 0$,
are:
\begin{align}
w_{+}^a =& 
\frac{1}{\sqrt{(\sqrt{ \gamma^2 - \omega^2} - \gamma_1)^2 + (\omega - \gamma_2 )^2}}
\begin{pmatrix}
\sqrt{ \gamma^2 - \omega^2} - \gamma_1 \\
- \gamma_2 + \omega
\end{pmatrix}
,
\\
w_{-}^a =& 
\frac{-1}{\sqrt{(\sqrt{ \gamma^2 - \omega^2} + \gamma_1)^2 + ( \omega - \gamma_2 )^2}}
\begin{pmatrix}
\sqrt{ \gamma^2 - \omega^2} + \gamma_1 \\
\gamma_2 - \omega
\end{pmatrix}
;
\end{align}
or, assuming $\gamma^2 \geqslant \omega^2$
\begin{align}
w_{+}^a =& 
\frac{1}{\sqrt{2} \sqrt{
		\gamma^2  -  \gamma_1\sqrt{ \gamma^2 - \omega^2} - \gamma_2 \omega 
}}
\begin{pmatrix}
\sqrt{ \gamma^2 - \omega^2} - \gamma_1 \\
- \gamma_2 + \omega
\end{pmatrix}
,
\\
w_{-}^a =& 
\frac{-1}{\sqrt{2} \sqrt{
		\gamma^2  +  \gamma_1\sqrt{ \gamma^2 - \omega^2} - \gamma_2 \omega 
}}
\begin{pmatrix}
\sqrt{ \gamma^2 - \omega^2} + \gamma_1 \\
\gamma_2 - \omega
\end{pmatrix}
.
\end{align}

The principal axis of the ellipses are related with the eigenvalues of the 
\emph{inverse optical matrix}.

If the unlensed angle $\delta \beta^a$ were $w_{+}^a$;
the one would see a lensed angle $\delta \theta^a$ given by
\begin{equation}
\delta \theta^a = \frac{1}{\lambda_+} w_{+}^a
,
\end{equation}
and if the unlensed angle $\delta \beta^a$ were $w_{-}^a$;
the one would see a lensed angle $\delta \theta^a$ given by
\begin{equation}
\delta \theta^a = \frac{1}{\lambda_-} w_{-}^a
;
\end{equation}
and since, for $\kappa \leqslant 1$, $\lambda_+ \geqslant |\lambda_-|$, one has that,
$w_{-}^a$ determines the direction
of the maximum deformation of the original image;
and in particular is the direction of the major axis of an elliptical
deformation when $\omega=0$.

The angle with the $x$ axis in the direction of the maximum deformation is then:
\begin{equation}
\tan(\phi_{xz}) = \frac{w_{-z}}{w_{-x}}
=
\frac{-\gamma_2+\omega}{-\sqrt{\gamma^2-\omega^2} - \gamma_1}
.
\end{equation}
And the vector pointing in the direction of the maximum deformation is
\begin{equation}
\pm \frac{1}{\lambda_-} w_{-}^a
;
\end{equation}
where $w_{-}^a$ is the normalized version.
While the vector pointing in the direction of the opposite deformation is
\begin{equation}
\pm \frac{1}{\lambda_+} w_{+}^a
;
\end{equation}
where $w_{+}^a$ is the normalized version.

Note that the eigenvectors are orthogonal only if $\omega=0$.
In what follows we assume $\omega=0$.

Note that the ratio of the components for the (-) eigenvector \eqref{eq:eigenv+menos+I}
\begin{equation}
\frac{-\sqrt{\gamma^2} -\gamma_1}{-\gamma_2}
=
\frac{-\sqrt{\gamma^2} -\gamma_1}{-\gamma_2}
\frac{-\sqrt{\gamma^2} +\gamma_1}{-\sqrt{\gamma^2} +\gamma_1}
=
\frac{-\gamma_2}{-\sqrt{\gamma^2} +\gamma_1}
,
\end{equation}
coincides with the ratio of the second prescription for the (-) eigenvector \eqref{eq:eigenv+menos+II};
so that we can use any of them.
\begin{figure}
	\centering
	\includegraphics[clip,width=0.15\textwidth]{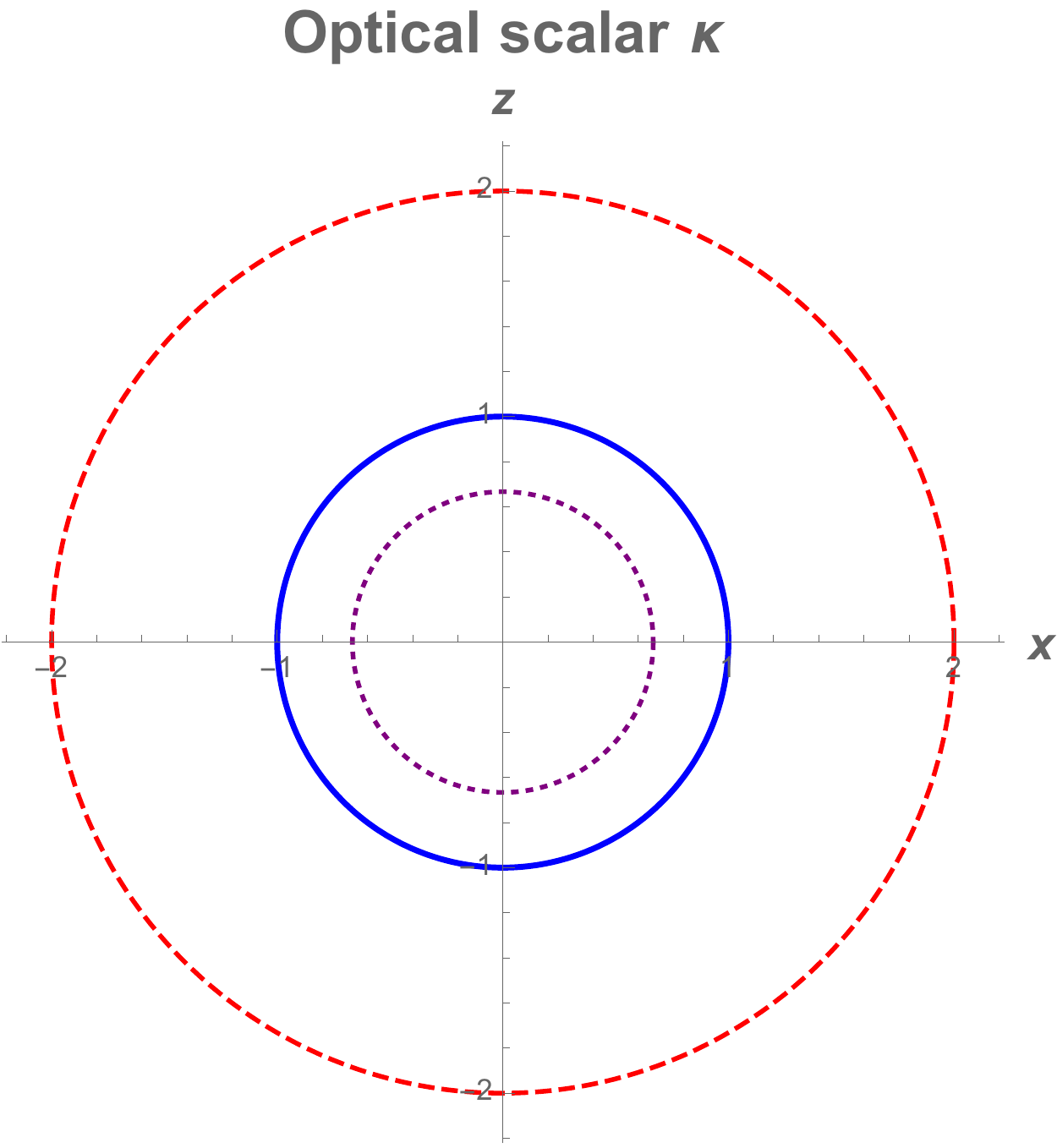}
	\includegraphics[clip,width=0.15\textwidth]{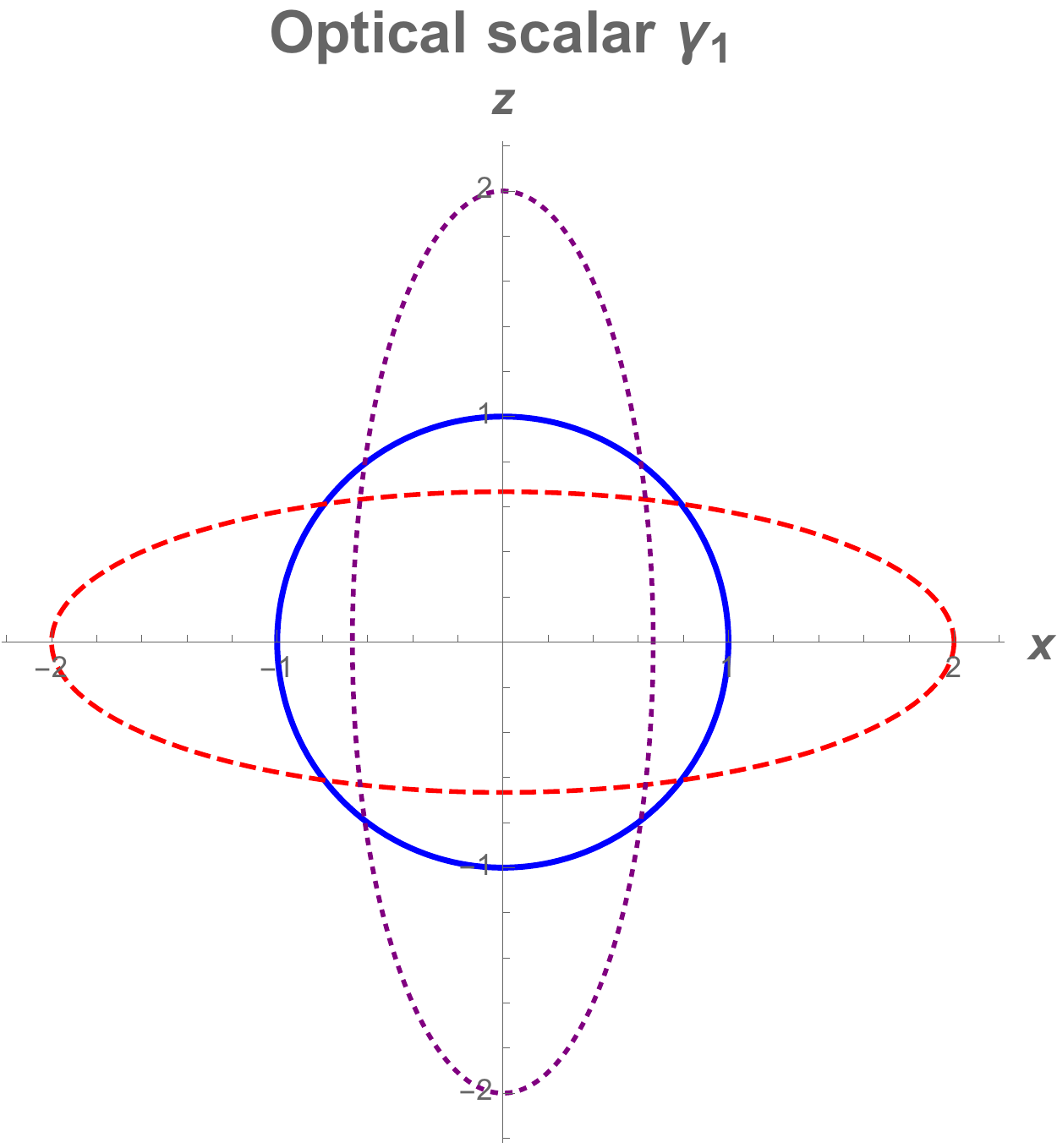}
	\includegraphics[clip,width=0.15\textwidth]{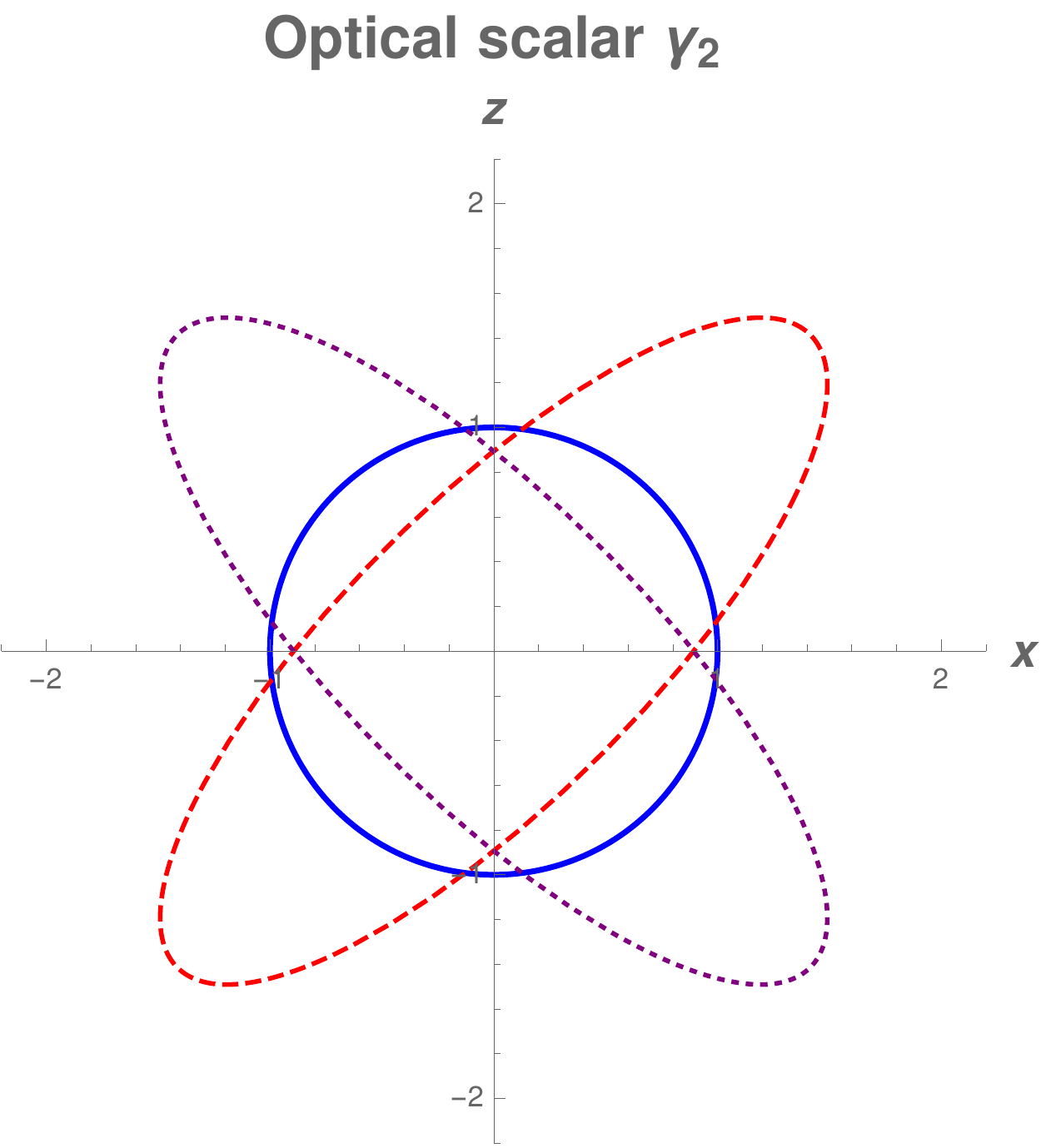}
	\caption{The isolated effect of the expansion and shear over a small circular 
		source (in continuous blue line).
		Positive values of the optical scalars produce distortions that are
		sketched in red and large dashed lines while the effect of negative values 
		is illustrated with small dashed lines in purple.
		 }\label{fig:KappaEffect}
\end{figure}

\section{Numerical calculation at fix angle near the Einstein ring}\label{sec:app-adiferentes}
Calculation along the exact geodesic for an angle close to the Einstein ring
shows very small variations of the optical scalar for different values
of the angular parameter.
For instance, at the angle $\alpha_x = 9.311007577025564404E-06$
we obtains
$\gamma_{1}= -1.934974633964382138E-01$ for $a=0$,
$\gamma_{1}= -1.934974633895238034E-01$ for $a=0.5$
and
$\gamma_{1}= -1.934974633822328950E-01$ for $a=0.98$.
At the other angular values we find also very small variations
of the optical scalar for different values of the angular
parameter.
	

\label{lastpage}
\end{document}